\documentclass[12pt]{article}
\usepackage[latin9]{inputenc}
\usepackage{geometry}
\usepackage{harvard}
\usepackage{amsthm}
\usepackage{amsmath}
\usepackage{amssymb}
\usepackage{graphicx}
\usepackage{setspace}
\usepackage{esint}
\usepackage{float}
\usepackage{subfig}
\usepackage{color}
\makeatletter

  \theoremstyle{definition}
  
  \theoremstyle{remark}
  \newtheorem{rem}{\protect\remarkname}
  \theoremstyle{plain}

\makeatother

  \providecommand{\definitionname}{Definition}
  \providecommand{\propositionname}{Proposition}
  \providecommand{\remarkname}{Remark}

\newtheorem{definition}{Definition}[section]

\def\b{\beta}

\def\s{\sigma}

\def\o{\omega}

\newcommand{\rr}{\mathbb{R}}
\newcommand{\pp}{\mathbb{P}}
\newcommand{\ff}{\mathcal{F}}

\newcommand{\media}{\mathbb{E}}

\chardef\o="1C

\begin{document}

\title{The Italian Pension Gap: \\ a Stochastic Optimal Control Approach}

\author{Alessandro Milazzo\footnote{Imperial College, London. Email: a.milazzo16@imperial.ac.uk. } \and Elena Vigna\footnote{Corresponding author. Universit\`{a} di Torino, Collegio Carlo Alberto and CeRP, Italy. Email: elena.vigna@unito.it. }}

\date{\today}

\maketitle
\thispagestyle{empty}

\begin{abstract}
We study the gap between the state pension provided by the Italian pension system pre-Dini reform and post-Dini reform. The goal is to fill the gap between the old and the new pension by joining a defined contribution pension scheme and adopting an optimal investment strategy that is target-based. We find that it is possible to cover, at least partially, this gap with the additional income of the pension scheme, especially in the presence of late retirement and in the presence of stagnant career. Workers with dynamic career and workers who retire early are those who are most penalised by the reform. Results are intuitive and in line with previous studies on the subject.\\

\textbf{Keywords.}  Pension reform, defined contribution pension scheme, net replacement ratio, stochastic optimal control, dynamic programming, HJB equation, Bellman's optimality principle.\\

\textbf{JEL classification:} C61, D81, G11.
\end{abstract}

\section{Introduction}
During the last few decades, several countries across the world have faced the ageing population problem. It is well known that the ageing of the population threatens the sustainability of Pays-As-You-Go pension systems, that is essentially based on a sufficiently high ratio workers to pensioner. To tackle this issue, many governments have introduced new reforms that lead to deep changes into the pension systems.\\
In particular, in Italy, the \textit{Dini reform} (introduced in 1995) instituted a completely different pension system for the new classes of workers. Before the reform, the public pension was provided through a defined benefit salary-related system. Namely, the pension provided was simply a service-based percentage of the last salary of the worker. After the reform, the public pension has been provided through a contribution-based system, where the worker contributes by himself to build his pension.\\
This remarkable change generated two different classes of workers, the pre-reform workers and the post-reform workers, and lead to a pension gap between their corresponding pension rates and replacement ratios.\\
The aim of this paper is to illustrate a possible way to fill this pension gap investing optimally in a defined contribution (DC) pension fund during the working life. In order to achieve the goal, a stochastic optimal control problem with suitable annual targets is solved. We consider two different salary growths to represent two different classes of workers: a linear salary growth (blue-collar workers) and an exponential salary growth (white-collar workers).\\
A numerical section illustrates the practical application of the model. Our results are in line with previous results on the comparison between the old pre-reform Italian pension and the new post-reform one, see \citeasnoun{borella-codamoscarola-gde} and \citeasnoun{borella-codamoscarola-jpef}. We find that the gap between salary-related pension and contribution-based pension is larger for workers with dynamic career than for workers with a stagnant career. A slow salary increase associated to late retirement can produce a new pension that is almost equal to (or even exceeds) the old pension. Expectedly, the gap is easier to cover in the case of late retirement, and vice versa. Interestingly, the gap increases when the rate of growth of the salary increases.\\
The reminder of the paper is as follows.
In Section 2 we introduce the milestones of the Italian pension system and the consequences of the Dini reform.
In Section 3 we build the model and the corresponding stochastic optimal control problem.
In Section 4 we derive the closed-form solutions to the problem for the two different salary growths considered.
In Section 5 we carry out some simulations in order to test the model and show the behaviour of the optimal investment strategy and the optimal fund growth in a base case scenario. In Section 6 we make sensitivity analysis of the pension distribution with respect to retirement age. In Section 7 we investigate the break even points that lead the ``new'' pension to equal the ``old'' pension. Section 8 concludes.

\section{The Italian pension provision}
The Italian pension system has been modified through a series of legislative measures taken by different governments during the 90s. We only hint at the \emph{Dini reform} which is useful in order to understand the following model.\\
The Dini reform (law 335, 1995) has changed the system for the calculation of the pension from a salary-based system to a contribution-related system. The workers shifted from one system to the other depending on the contributions paid at the end of 1995. Therefore, three different situations were created:
\begin{enumerate}
\item The workers with at least eighteen years of contributions on 31/12/1995 remained under the salary-related system and therefore they were not touched by the reform.
\item The workers with less than eighteen years of contributions on 31/12/1995 were subjected to a mixed method.
\item The workers who were first employed after 31/12/1995 are subjected to the contribution-based system.
\end{enumerate}
In this paper, we compare the method for the calculation of the public pension before the Dini reform with the ``new'' method for the calculation of the public pension in Italy after the reform.\\
The formula for the pension rate $P_o$ before the Dini reform was
\begin{equation} \label{oldpension}
P_o = 0.02\cdot T\cdot S(T),
\end{equation}
where $S(T)$ is the final salary and $T$ indicates the number of past working years. In the following, we shall call the pension rate \eqref{oldpension} the ``old pension''. The old pension is a percentage of the product of the last salary by the years of service, and the related net replacement ratio --- which is the ratio between the first pension rate received after retirement and the last salary perceived before retirement --- is given by
$$
\Pi_o = \frac{P_o}{S(T)} = 0.02\cdot T.
$$
In contrast, the new formula for the pension rate $P_n$ is described by
\begin{equation} \label{newpension}
P_n=\beta\cdot c\cdot \sum_{t=0}^{T-1}S(t)(1+w)^{T-t},
\end{equation}
where
\begin{itemize}
\item $\beta$ is the conversion coefficient between a lump sum and the annuity rate, and its choice should reflect actuarial fairness. If it does, then $\beta=1/\ddot{a}_x$, where $\ddot{a}_x$ is the single premium of a lifetime annuity issued to a policyholder aged $x$, i.e.,
    \begin{equation}\label{annuity}  \ddot{a}_x= \sum_{n=1}^{\omega-x} \, _np_x \cdot v^n, \end{equation}
where $\omega$ is the extreme age, $v=1/(1+r)$ is the annual discount factor and $_np_x$ is the survival probability from age $x$ to age $x+n$.
\item $c$ is the contribution percentage for the calculation of the pension rate (supposed to be constant during the whole working life and, for the employees, set by law to 33\%).
\item $T$ indicates the number of past working years.
\item $w$ is the mean real GDP increase.
\end{itemize}
In the following, we shall call the pension rate \eqref{newpension} the ``new pension''. Compared to the old pension, the new pension is given by a complicated formula and depends not only on all the salaries but also on other parameters. Its related net replacement ratio is
$$
\Pi_n = \frac{P_n}{S(T)}.
$$

\section{The optimization problem}
The main idea of this paper is the following.\\
\indent We assume that the worker was firstly employed after 31/12/1995 and will receive the new pension \eqref{newpension} from the first pillar (i.e., the public pension), but he wants to integrate it with additional income from the second pillar (i.e., the private pension funds) to obtain a pension rate that is as close as possible to the one that he would have obtained with the old pension rule \eqref{oldpension}. Since the Dini reform, and apart from a few exceptions accessible only by self-employed (not considered in this paper), pension funds in Italy are defined contribution (DC) and not defined benefit (DB). This means that the contribution to be paid into the fund is fixed a priori in the scheme's rules and the benefit obtained at retirement depends on the investment performance of the fund in the accumulation period.\\
\indent We assume that at time 0 the worker joins a DC pension scheme and has control on the investment strategy to be adopted on the time horizon $[0,T]$, where $T$ is the retirement time. The financial market consists of two assets, a riskless asset $B=\{B(t)\}_{t\geq 0}$ and a risky asset $Z=\{Z(t)\}_{t\geq 0}$, whose dynamics are described by
\begin{equation} \label{risklessasset}
dB(t) = rB(t)dt,
\end{equation}
\begin{equation} \label{riskyasset}
dZ(t) = \mu Z(t)dt + \sigma Z(t)dW(t),
\end{equation}
where $r$ is a constant rate of interest and $\{W(t)\}_{t\geq 0}$ is a standard Brownian motion defined and adapted on a complete filtered probability space $(\Omega,\ff,\{\ff_t\}_{t\geq 0},\pp)$. We assume that the contribution $c(t)$ paid into the fund at time $t$ is a fixed proportion of the salary of the member
$$c(t)=kS(t), \quad t\in[0,T],$$
where $k\in(0,1)$ and $S(t)$ is the salary of the member at time $t$. Finally, the proportion of portfolio invested into the risky asset at time $t\in[0,T]$ is $y(t)$. Hence, the dynamics of the wealth are described by the following SDE
\begin{equation} \label{fundSDE}
\left\{
\begin{array}{ll}
dX(t) = \{[(\mu-r)y(t)+r]X(t)+c(t)\}dt+\sigma y(t)X(t)dW(t)\\
X(0)=x_0
\end{array}
\right.
\end{equation}
where $x_0 \geq 0$ is the initial wealth paid into the fund (it can also be a transfer value from another pension fund).
Because the aim of the worker is to reach a pension rate that is as close as possible to that of the salary-related method, we assume that there exist annual targets $\{F(t)\}_{t=0,1,2,...T}$ that he wants to achieve, and that his preferences are described by the loss suffered when the targets are not met. Thus, we introduce the following quadratic loss (or disutility) function
$$L(t,X(t)) = (F(t)-X(t))^2, \quad t\in[0,T].$$
\begin{rem}
The use of a quadratic loss is very common in the context of pension funds. Some examples are \citeasnoun{boulier-trussant-florens}, \citeasnoun{boulier-michel-wisnia}, \citeasnoun{cairns}, \citeasnoun{gerrard-haberman-vigna}, \citeasnoun{gerrard-haberman-vigna06}, \citeasnoun{gerrard-hoejgaard-vigna}. Moreover, \citeasnoun{vigna-qf} and \citeasnoun{menoncin-vigna} deeply analyse and discuss the link between ``utility-based'' approach and ``target-based'' approach. From a theoretical point of view, the quadratic loss function also penalizes any deviations above the target, and this can be considered as a drawback to the model. However, the choice of trying to achieve a target and no more than this has the effect of a natural limitation on the overall risk of the portfolio: once the desired target is reached, there is no reason for further exposure to risk and therefore the surplus becomes undesirable. This is in accordance also with the fact that the mean-variance approach to portfolio selection has been shown to be equivalent to the minimization of a quadratic loss function: see the seminal papers by \citeasnoun{zhou-li} and \citeasnoun{li-ng}, and, in the context of DC pension schemes, \citeasnoun{vigna-qf}. The idea that people act by following subjective targets is also accepted in the decision theory literature. For instance, \citeasnoun{kahneman-tversky} support the use of targets in the cost function, and \citeasnoun{bordley-licalzi} investigate and support the target-based approach in decision making under uncertainty.\end{rem}

%
We now need to define the targets.
Recalling that the worker's goal is to reach the old pension pre-Dini reform, we set as final target $F(T)$ the amount that the retiree aged $x$ should pay to an insurance company in order to fill the gap between the previous pension rate and the current one, i.e.,
\begin{equation} \label{finaltarget}
F(T) = (P_o-P_n)\ddot{a}_x ,
\end{equation}
where $\ddot{a}_x$ is the price of the annuity given by \eqref{annuity} to a retiree aged $x$, $P_o$ is given by $(\ref{oldpension})$, $P_n$ is the continuous formulation of equation $(\ref{newpension})$, which means
$$P_n=\beta c \int_0^T S(t)e^{w(T-t)}dt.$$

The interim targets $F(t)$ for $t\in [0,T)$ are set to be the compounded value of fund plus contributions using the interest rate $r^*$ that matches a continuity condition between interim targets and final target, i.e.
\begin{equation}\label{compoundedtargets}
F(t)=x_0e^{r^*t}+\int_0^t c(s)e^{r^*(t-s)}ds,
\end{equation} with $r^*$ such that\footnote{We will approximate the value $r^*$ with Newton-Raphson algorithm.}
\begin{equation} \label{continuitycond}
\lim_{t\rightarrow T^-} F(t)=F(T).
\end{equation}

The worker's goal is to minimize the conditional expected losses that can be experienced from the fund until retirement
\begin{equation} \media_{0,x_0}\left[\int_0^T e^{-\rho s}L(s,X(s))ds+e^{-\rho T}L(T,X(T))\right],\end{equation}
where $\rho>0$ is the (subjective) intertemporal discount factor.\\
As usual in optimization problems in DC pension schemes, the contribution rate is \emph{not} a control variable, and the only control variable for the worker is the share of portfolio $y(t)$ to be invested into the risky asset at time $t\in[0,T]$. To formulate the optimization problem, we define the performance criterion at time $t$ with wealth $x$, i.e.,
\begin{equation} \label{Jform}
J_{t,x}(y(\cdot))=\media_{t,x}\left[\int_t^T e^{-\rho s}L(s,X(s))ds+e^{-\rho T}L(T,X(T))\right]
\end{equation}
and the admissible strategies.
\begin{definition}
An investment strategy $y(\cdot)$ is said to be \textbf{admissible} if
$y(\cdot)\in L^2_\mathcal{F}(0,T;\rr)$.
\end{definition}

\noindent The minimization problem, then, becomes
\begin{equation}\label{min-prob}\mbox{Minimize  } J_{0,x_0}(y(\cdot))\end{equation}
over the set of admissible strategies.

\section{Solution}
In order to solve the optimization problem \eqref{min-prob}, the value function is defined as
\begin{equation} \label{Vform}
V(t,x):=\inf_{y(\cdot)} J_{t,x}(y(\cdot)) \quad \forall(t,x)\in U=[0,T]\times(-\infty,+\infty).
\end{equation}

\begin{rem}
In this work, we neither set boundaries on the values that the fund $X(\cdot)$ can assume, nor set boundaries on the share $y(t)$ to be invested in the risky asset. The existence of a minimum finite bound on $X(t)$ would be desirable, as well as proper boundaries on the investment strategy. The former would be intended to protect the retiree from outliving his asset and not being able to buy a minimum level of pension at time $T$, the latter to comply with the usual forbiddance of short-selling and borrowing. However, adding restrictions to the state variable and the control variable means adding boundary conditions to the problem and this makes it extremely hard (and often impossible) to solve analytically. Among the few works that treat optimization problems with restrictions in DC pension schemes, see \citeasnoun{digiacinto-gozzi-federico-08} and \citeasnoun{digiacinto-federico-gozzi-vigna-ejor}.\end{rem}

We write the HJB equation
\begin{equation} \label{HJBmodel}
\inf_{y\in\rr}[e^{-\rho t}L(t,x)+\mathcal{L}^yV(t,x)]=0, \quad \forall \; (t,x)\in U
\end{equation}
$$V(T,x)=e^{-\rho T}L(T,x), \quad \forall \; x\in\rr,$$
where
$$\mathcal{L}^uf(t,x)=\frac{\partial}{\partial t}f(t,x)+b(t,x,u)\frac{\partial}{\partial x}f(t,x)+\frac 1 2 \sigma^2(t,x,u)\frac{\partial^2}{\partial x^2}f(t,x)$$
is the infinitesimal operator and the functions $b(\cdot)$ and $\sigma(\cdot)$ are the drift and the diffusion terms of the process $X=\{X(t)\}_{t\geq 0}$ defined by $(\ref{fundSDE})$.\\
Substituting into $(\ref{HJBmodel})$, we obtain $\forall(t,x)\in U$,
\begin{equation} \label{explicitHJB}
\inf_{y\in\rr}\left\{e^{-\rho t}(F(t)-x)^2+\frac{\partial V}{\partial t}+[x(y(\mu-r)+r)+c(t)]\frac{\partial V}{\partial x} +\frac 1 2 x^2y^2\sigma^2\frac{\partial^2 V}{\partial x^2}\right\}=0,
\end{equation}
with the boundary condition
\begin{equation} \label{boundarycond}
V(T,x)=e^{-\rho T}L(T,x).
\end{equation}
To have an easier notation, let us define
\begin{equation} \label{psiform}
\psi(t,x,y):=e^{-\rho t}(F(t)-x)^2+\frac{\partial V}{\partial t}+[x(y(\mu-r)+r)+c(t)]\frac{\partial V}{\partial x} +\frac 1 2 x^2y^2\sigma^2\frac{\partial^2 V}{\partial x^2}.
\end{equation}
Thus, equation $(\ref{explicitHJB})$ becomes
\begin{equation} \label{psiequation}
\inf_{y\in\rr}\psi(t,x,y)=0 \quad \Rightarrow \quad \psi(t,x,y^*)=0.
\end{equation}
The first and second order conditions are
\begin{equation} \label{firstorder}
\psi_y(t,x,y^*)=0,
\end{equation}
\begin{equation} \label{secondorder}
\psi_{yy}(t,x,y^*)>0.
\end{equation}
Therefore, $(\ref{firstorder})$ becomes
$$x(\mu-r)\frac{\partial V}{\partial x}+x^2y^*\sigma^2\frac{\partial^2 V}{\partial x^2}=0,$$
so that
\begin{equation} \label{opimtcontrol}
y^*=-\frac{\mu-r}{\sigma}\frac{1}{x\sigma}\frac{V_x}{V_{xx}}.
\end{equation}
Moreover, condition $(\ref{secondorder})$ is satisfied if and only if
\begin{equation} \label{necessarycondition}
x^2\sigma^2\frac{\partial^2 V}{\partial x^2}>0 \quad \Leftrightarrow \quad \frac{\partial^2 V}{\partial x^2}>0.
\end{equation}
We will show later that this condition is actually satisfied, so that the solution is a minimum.\\
By substituting $(\ref{opimtcontrol})$ into $(\ref{psiequation})$, we obtain the non-linear PDE
\begin{equation} \label{finalHJB}
e^{-\rho t}(F(t)-x)^2+V_t+[rx+c(t)]V_x-\frac 1 2 \left(\frac{\mu-r}{\sigma}\right)^2\frac{V_x^2}{V_{xx}}=0.
\end{equation}
We guess a solution of the form
\begin{equation} \label{Vguess}
V(t,x)=e^{-\rho t}[\alpha(t)x^2+\beta(t)x+\gamma(t)].
\end{equation}
From the boundary condition $(\ref{boundarycond})$, we obtain
$$e^{-\rho T}(F(T)-x)^2=e^{-\rho T}[\alpha(T)x^2+\beta(T)x+\gamma(T)], \quad \forall x\in(-\infty,+\infty),$$
so that
\begin{equation} \label{boundaryconds}
\alpha(T) = 1, \quad \beta(T)=-2F(T), \quad \gamma(T)=[F(T)]^2.
\end{equation}
The partial derivatives of $V$ are
$$V_t(t,x) = -\rho e^{-\rho t}[\alpha(t)x^2+\beta(t)x+\gamma(t)]+e^{-\rho t}[\alpha '(t)x^2+\beta '(t)x+\gamma '(t)],$$
$$V_x(t,x) = e^{-\rho t}[2\alpha(t)x+\beta(t)], \quad V_{xx}(t,x)=2e^{-\rho t}\alpha(t),$$
and substituting them into $(\ref{opimtcontrol})$, we derive the \emph{optimal investment strategy} at time $t$ with wealth $x$, i.e.,
\begin{equation} \label{optinvestment}
y^*(t,x) = -\frac{\mu-r}{\sigma}\frac{1}{x\sigma}\left(x+\frac{\beta(t)}{2\alpha(t)}\right).
\end{equation}
Substituting the partial derivatives of $V(\cdot,\cdot)$ into $(\ref{finalHJB})$, we have
\begin{align} \label{finalHJB2}
&[1-\rho\alpha(t)+\alpha '(t)+2r\alpha(t)-\lambda^2\alpha(t)]x^2+[-2F(t)-\rho\beta(t)+\beta '(t)+r\beta(t)+\nonumber\\
&+2\alpha(t)c(t)-\lambda^2\beta(t)]x+\left[F(t)^2-\rho\gamma(t)+\gamma '(t)+c(t)\beta(t)-\lambda^2\frac{\beta(t)^2}{4\alpha(t)}\right]=0,
\end{align}
where $\lambda:=(\mu-r)/\sigma$ is the \emph{Sharpe ratio} of the risky asset.\\
Since $(\ref{finalHJB2})$ must hold $\forall(t,x)\in U$, we derive the following system of ODE's
\begin{equation} \label{systemODE}
\left\{
\begin{array}{lll}
\alpha '(t)=[\rho+\lambda^2-2r]\alpha(t)-1=a\alpha(t)-1\\
\beta '(t)= [\rho+\lambda^2-r]\beta(t)+2F(t)-2c(t)\alpha(t)=\tilde{a}\beta(t)+2F(t)-2c(t)\alpha(t)\\
\gamma '(t)= \rho\gamma(t)-F(t)^2-c(t)\beta(t)+\lambda^2\frac{\beta(t)^2}{4\alpha(t)}
\end{array}
\right.
\end{equation}
where we have defined $a:=\rho+\lambda^2-2r$, $\tilde{a}=a+r$, and with the boundary conditions $(\ref{boundaryconds})$.

\subsection{Solution of the problem with two different salary evolutions}\label{sec:solution-with-salaries}

Two different salaries are compared: a linear salary
\begin{equation} \label{lsalary}
S_l(t) = S_0(1+g_l t), \quad t\in[0,T],
\end{equation}
and an exponential salary
\begin{equation} \label{esalary}
S_e(t) = S_0e^{g_e t}, \quad t\in[0,T],
\end{equation}
where $S_0$ is the initial salary and $g_i$ ($i=l,e$) is the mean real salary increase.

The contribution in the two cases is
$$c_i(t)=k_i S_i(t), \quad t\in[0,T], \; i=l,e,$$
\begin{rem}
We have selected two simple models for the salary growth for analytical tractability and the aim of providing closed-form solutions for the optimal investment strategy.\footnote{ For a more accurate and realistic model for the salary growth in the Italian context we refer to the micro-simulation model developed by \citeasnoun{borella-codamoscarola-gde} and \citeasnoun{borella-codamoscarola-jpef}.} The two different salary growths may represent two different categories of workers, the exponential growth being associated to white-collar workers with dynamic salary increase, the linear salary increase being associated to blue-collar workers with smooth salary increase. The distinct $k's$ reflect the assumption that the savings capacity of white-collar workers is higher than the savings capacity of blue-collar workers.
\end{rem}
Therefore, there are also two different families of targets.\\
For the linear salary case (for notational convenience, in the following we will write $g$ and $k$ in the place of $g_l$ and $k_l$):
\begin{align} \label{linintertarg}
F_l(t) &= x_0e^{r^*t}+\int_0^t kS_0(1+gs)e^{r^*(t-s)}ds \\ \nonumber
&=\left[x_0+\frac{kS_0}{r^*}+\frac{kgS_0}{(r^*)^2}\right]e^{r^*t}- \frac{kgS_0}{r^*}t-\frac{kS_0}{r^*} \left[1+\frac{g}{r^*}\right],
\end{align}
\begin{equation} \label{linfintarg}
F_l(T)=(\Pi_o-\Pi_n^l)S_l(T)\ddot{a}_x,
\end{equation}
where $\Pi_n^l$ is the net replacement ratio for the new public pension with linear salary.\\
For the exponential salary case (for notational convenience, in the following we will write $g$ and $k$ in the place of $g_e$ and $k_e$):
\begin{equation} \label{expintertarg}
F_e(t) = x_0e^{r^*t}+\int_0^t kS_0e^{gs}e^{r^*(t-s)}ds = \left[x_0-\frac{kS_0}{g-r^*}\right]e^{r^*t}+\frac{kS_0}{g-r^*}e^{gt}.
\end{equation}
\begin{equation} \label{expfintarg}
F_e(T)=(\Pi_o-\Pi_n^e)S_e(T)\ddot{a}_x,
\end{equation}
where $\Pi_n^e$ is the net replacement ratio for the new public pension with exponential salary.\\
Solving the system $(\ref{systemODE})$ in both cases, we find the following solutions
\begin{equation} \label{alphaform}
\alpha(t)=\left(1-\frac{1}{a}\right)e^{-a(T-t)}+\frac{1}{a},
\end{equation}
\begin{align*}
\beta_l(t) &= -2F_l(T)e^{-\tilde{a}(T-t)}+\frac{k_4}{r}\left(1+\frac{g}{r}\right)e^{(r-\tilde{a})(t-t)} +\frac{1}{\tilde{a}}\left(k_5+\frac{k_1g}{a\tilde{a}}+\frac{k_2g}{\tilde{a}}\right)+\\
&+\frac{k_4g}{r}te^{(r-\tilde{a})(T-t)}+\frac{g}{\tilde{a}}\left(\frac{k_1}{a}+k_2\right)t+ \frac{k_3}{r^*-\tilde{a}}\left[e^{r^*t}-e^{\tilde{a}t+(r^*-\tilde{a})T}\right]+\\
&-\left[\frac{k_4}{r}+\frac{k_5}{\tilde{a}}+\frac{k_4g}{r^2}+\frac{k_1g}{a(\tilde{a})^2}+ \frac{k_2g}{(\tilde{a})^2}+\left(\frac{k_4g}{r}+\frac{k_1g}{a\tilde{a}}+ \frac{k_2g}{\tilde{a}}\right)T\right]e^{-\tilde{a}(T-t)},
\end{align*}
\begin{align*}
\beta_e(t) = &- 2F_e(T)e^{-\tilde{a}(T-t)} -\frac{k_4}{g-r}e^{(g+a)t-aT}- \frac{\tilde{k_4}}{g-\tilde{a}}e^{gt}+\\
&+\left(\frac{k_4}{g-r}+\frac{\tilde{k_4}}{g-\tilde{a}}\right)e^{\tilde{a}t+(g-\tilde{a})T}- \frac{\tilde{k_3}}{r^*-\tilde{a}}\left[e^{\tilde{a}t+(r^*-\tilde{a})T}-e^{r^*t}\right],
\end{align*}
\begin{align*}
\gamma_i(t) = &+ e^{-\rho(T-t)}\int_t^T\left[F_i(s)^2+c_i(s)\beta_i(s)+ \lambda^2\frac{\beta_i(s)^2}{4\alpha(s)}\right]e^{\rho(T-s)}ds+\\
&+F_i(T)^2e^{-\rho(T-t)} \quad \qquad i=l,e
\end{align*}
where
\begin{align*}
\tilde{a}&=a+r, \quad k_1=2kS_0, \quad k_2=\frac{2kS_0}{r^*}, \quad k_3=2x_0+k_2+\frac{2kgS_0}{(r^*)^2}, \quad k_4=k_1-\frac{k_1}{a},\\ k_5&=\frac{k_1}{a}+k_2\left(1+\frac{g}{r^*}\right), \quad \tilde{k_2}=\frac{2kS_0}{g-r^*}, \quad \tilde{k_3}=2x_0-\tilde{k_2}, \quad \tilde{k_4}=\frac{k_1}{a}-\tilde{k_2}.
\end{align*}
Substituting these solutions into $(\ref{optinvestment})$, we obtain the two optimal investments
$$y_l^*(t) \quad \mbox{and} \quad y_e^*(t), \quad t\in[0,T].$$
Finally, we observe that condition $(\ref{necessarycondition})$ is satisfied. Indeed,
$$V_{xx}=2e^{-\rho t}\alpha(t)=2e^{-\rho t}(e^{-a(T-t)}+a^{-1}(1-e^{-a(T-t)})).$$
If $a>0$, then $(e^{-a(T-t)}+a^{-1}(1-e^{-a(T-t)}))>0$, obviously.\\
If $a<0$, then $(e^{-a(T-t)}+a^{-1}(1-e^{-a(T-t)}))>0$, because $a^{-1}<0$ and also $1-e^{-a(T-t)}<0$.

\section{Simulations }
We have carried out several numerical simulations in order to investigate some quantities of interest to the pension fund member when the model is implemented in the practice. In particular, we have investigated to what extent the gap between the old pension and the new pension is filled.
We have first considered a base case, and then made some sensitivity analysis.

\subsection{Base case: Assumptions}\label{sec:base-case-assumptions}

The assumptions for the base case are the following:
\begin{itemize}
\item the initial fund is $X(0)=1$;
\item the public pension contribution is $c=33\%$;
\item the mean GDP growth rate is $w=1.5\%$;
\item the riskless interest rate is $r=1.5\%$;
\item the drift of the risky asset is $\mu=6\%$;
\item the diffusion of the risky asset is $\s=12\%$;
\item the intertemporal discount factor is $\rho=3\%$;
\item the annuity value is calculated with the Italian projected mortality table IP55 (for males born between 1948 and 1960); it is $\ddot{a}_{65}(1.5\%)=17.875$; therefore, the conversion factor from lump sum to annuity is $\b=1/\ddot{a}_{65}(1.5\%)=0.0056$;
\item the age when the member joins the scheme is $x_0=30$;
\item the time horizon is $T=35$ meaning that the retirement age is $x_T=65$;
\item the initial salary is $S(0)=1$.
\end{itemize}
We have considered two different salary growths, with different values for the annual salary increase $g$ and the annual contribution rate $k$:
\begin{itemize}
\item exponential salary: $S_{e}(\cdot)$ as in formula $(\ref{esalary})$ with $g_{e}=6\%$ and salary contribution percentage $k_{e}=10\%$;
\item linear salary: $S_{l}(\cdot)$ as in formula $(\ref{lsalary})$ and $g_{l}=8\%$ and salary contribution percentage $k_{l}=4\%$.
\end{itemize}

In the simulations, we have discretized the Brownian motion with a discretization step equal to two weeks ($\Delta t=1/26$), and we have simulated its behaviour over time in 1000 different scenarios. At each time point, we have not applied the optimal unconstrained investment strategy $y^*(t)$, because short-selling and borrowing are likely to be forbidden. We have, instead, implemented the ``sub-optimal'' constrained investment strategies, which are constrained to stay between 0 and 1. In particular, the sub-optimal $y^{so}(t)$ is defined as
\begin{equation*} y^{so}(t)=\left\{\begin{array}{ll}
0 & \mbox{for } y^*(t)<0 \\
 y^*(t) & \mbox{for } y^*(t)\in[0,1]\\
 1 & \mbox{for } y^*(t)>0
 \end{array} \right.\end{equation*} where $y^*(t)$ is the optimal investment strategy.\footnote{In the following figures we have denoted the ``sub-optimal'' constrained investment strategy by $y^*(\cdot)$ and the ``sub-optimal'' constrained fund growth by $X^*(\cdot)$ for the sake of simplicity.}\\
The adoption of constrained investment strategies leads, as a desirable consequence, the fund not to run under $0$.
Constrained suboptimal strategies of this type are not new in the literature and they
are good approximations of the optimal investment strategies. They were applied, e.g., by \citeasnoun{gerrard-haberman-vigna06} and \citeasnoun{vigna-qf}
in the context of DC pension schemes with a constant interest rate,
and they proved to be satisfactory: with respect to the unrestricted
case the effect on the final results turned out to be negligible and
the controls resulted to be more stable over time.
In each scenario of market returns, the sub-optimal value $y^{so}(t)$, $t\in[0,T]$, has been calculated and then adopted for the fund growth.\\

\subsection{Base case: Results}\label{sec:base-case-results}
\begin{figure}[H]
		\centering
		\includegraphics[scale=0.4]{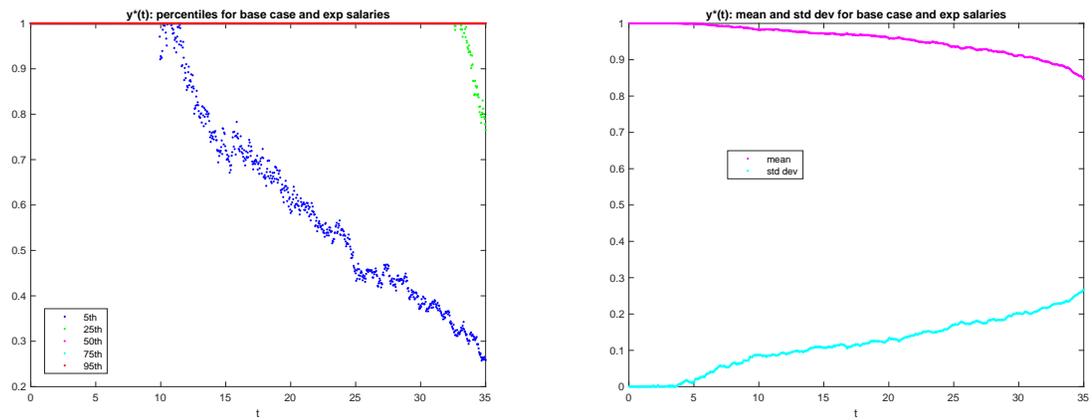}
		\caption{Optimal investment strategy, exponential salary growth}
		\label{fig-y-new-exp}
	\end{figure}

\begin{figure}[H]
		\centering
		\includegraphics[scale=0.4]{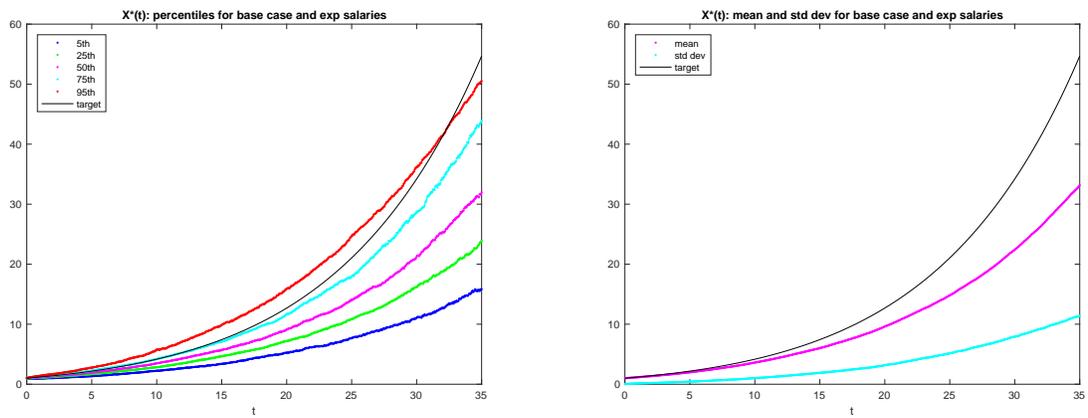}
		\caption{Fund growth, exponential salary growth}
		\label{fig-X-new-exp}
	\end{figure}

\begin{figure}[H]
		\centering
		\includegraphics[scale=0.4]{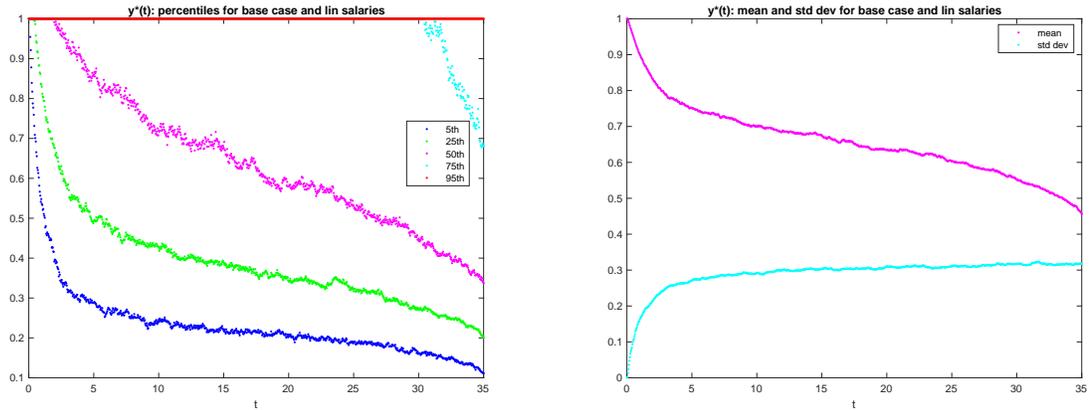}
		\caption{Optimal investment strategy, linear salary growth}
		\label{fig-y-new-lin}
	\end{figure}

\begin{figure}[H]
		\centering
		\includegraphics[scale=0.4]{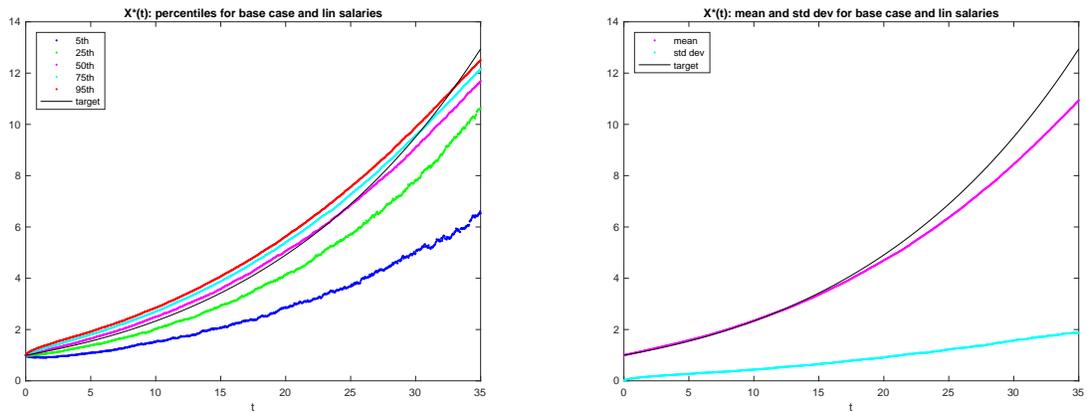}
		\caption{Fund growth, linear salary growth}
		\label{fig-X-new-lin}
	\end{figure}

Regarding the investment strategy adopted and the evolution of the fund over time in the base case, we present the following results:
\begin{itemize}
\item Table 1 reports, for both cases of exponential and linear growths, the old pension $P_o$, the new pension $P_n$, the old net replacement ratio $\Pi_o$, the new replacement ratio $\Pi_n$, the rate of growth of the targets $r^*$ and the last salary $S(T)$;

\item the two graphs of Figure \ref{fig-y-new-exp} report some percentiles (graph on the left) and mean and standard deviation (graph on the right) of the distribution, over the 1000 scenarios, of the investment strategy $y^{so}(t)$ for $t\in[0,T]$, for the exponential salary growth;

    \item the two graphs of Figure \ref{fig-X-new-exp} report some percentiles (graph on the left) and mean and standard deviation (graph on the right) of the distribution, over the 1000 scenarios, of the fund $X^{so}(t)$ for $t\in[0,T]$, for the exponential salary growth;

    \item the two graphs of Figure \ref{fig-y-new-lin} report some percentiles (graph on the left) and mean and standard deviation (graph on the right) of the distribution, over the 1000 scenarios, of the investment strategy $y^{so}(t)$ for $t\in[0,T]$, for the linear salary growth;

\item the two graphs of Figure \ref{fig-X-new-lin} report some percentiles (graph on the left) and mean and standard deviation (graph on the right) of the distribution, over the 1000 scenarios, of the fund $X^{so}(t)$ for $t\in[0,T]$, for the linear salary growth.
\end{itemize}

\begin{center}
\begin{tabular}{lllllll}
\hline
 & $\mathbf{P_o}$ & $\mathbf{P_n}$ & $\mathbf{\Pi_o}$ & $\mathbf{\Pi_n}$ & $\mathbf{r^*}$ & $\mathbf{S(T)}$\\ \hline
$\mathbf{S_e}$ & 5.716 & 2.657 & 0.7 & 0.325 & 0.078&8.166\\ \hline
$\mathbf{S_l}$ & 2.66 & 1.936 & 0.7 & 0.509 & 0.049&3.8\\ \hline
\end{tabular}\vspace{0.3cm}\\

Table 1.
\end{center}


We observe the following:
\begin{enumerate}
\item as expected, with exponential salary growth the final salary is significantly larger (more than double) than that with linear salary growth; therefore, despite the old replacement ratio (that does not depend on the salary growth) is the same ($70\%$), the old pension is much larger with a dynamic career than with a smooth one;
\item the investment strategy for the exponential growth is remarkably riskier than that for the linear growth: for the exponential growth in almost $75\%$ of the cases the portfolio is entirely invested in the risky asset for all $t$, while for the linear growth all percentiles of $y^{so}(t)$ (apart from the $95th$ one) decrease gradually from 1 to 0 over time;
\item the larger riskiness of the strategy for the exponential growth is due to the larger gap between the old and the new pension: the old replacement ratio is $70\%$ in both cases, but the new replacement ratio (that does depend on the salary growth) is $32\%$ for the exponential growth and $51\%$ for the linear growth; the larger gap to fill in for the exponential increase entails riskier strategies, and the smaller gap to fill in for the linear increase entails less risky strategies; these results seem to suggest that the new reform affects to a larger extent workers with a dynamic career than workers with a smooth career;
\item with both salary increases, on average the investment in the risky asset decreases over time and approaches 0 when retirement approaches; this result is in line with previous results on optimal investment strategies for DC pension schemes (see e.g. \citeasnoun{haberman-vigna}) and is consistent with the lifestyle strategy (see \citeasnoun{cairns}), that is an investment strategy largely adopted in DC pension funds in the UK.

\end{enumerate}

Finally, because the aim of this work is to reduce the gap between the old and the new pension, it is fundamental to investigate to what extent the gap is reduced. If the worker joins the pension fund for $T$ years, then at retirement he will receive the new pension $P_n$ plus the additional pension $P_{add}$ provided by the pension fund. Therefore, his total pension will be $P_{tot}$, given by
\[ P_{tot}=P_n+P_{add}\]
where
\[ P_{add}=\frac{X^{so}(T)}{\ddot{a}_{65}(1.5\%)}. \]

\begin{figure}[H]
		\centering
		\includegraphics[scale=0.4]{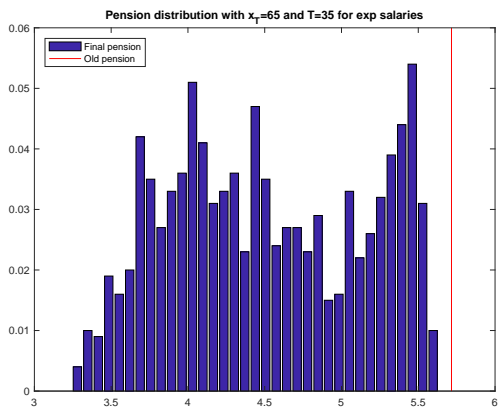}
		\caption{Distribution of final pension, exponential salary growth}
		\label{fig-pens-new-exp}
	\end{figure}

\begin{figure}[H]
		\centering
		\includegraphics[scale=0.4]{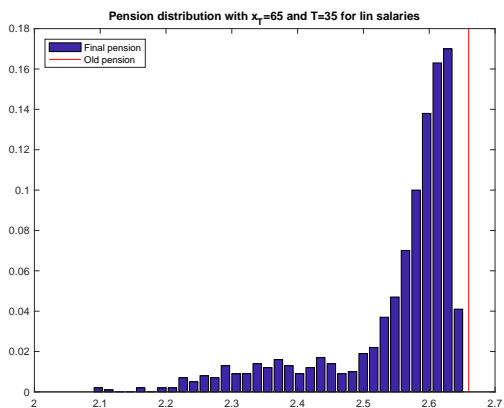}
		\caption{Distribution of final pension, linear salary growth}
		\label{fig-pens-new-lin}
	\end{figure}

In order to compare the old pension with the total pension,
\begin{itemize}
\item Figure \ref{fig-pens-new-exp} reports, in the case of exponential salary growth, the distribution, over the 1000 scenarios, of the final pension $P_{tot}$ that the retiree will receive; the old pension $P_o$ is also indicated, as a benchmark;
    \item Figure \ref{fig-pens-new-lin} reports, in the case of linear salary growth, the distribution, over the 1000 scenarios, of the final pension $P_{tot}$ that the retiree will receive; the old pension $P_o$ is also indicated, as a benchmark.
\end{itemize}

We notice that
\begin{itemize}
\item in the case of exponential salary growth, the final pension is distributed more uniformly between 3.3 and 5.6 (the old pension being 5.7), while for the linear salary growth there is a large concentration of the final pension on the immediate left of the target, between 2.5 and the target 2.66;
    \item what observed above is due to the fact that it is easier to reach the target in the case of linear increase than in the case of exponential increase, because (as observed in point 3. above) the gap between the old and the new pension is larger with the exponential increase than with the linear increase;
        \item the fact that it is relatively easy to approach the target with the linear increase is consistent with the rate of increase of the annual targets $r^*=4.86\%$ (see Table 1). This rate lies between the return on the riskless asset ($1.5\%$) and the expected return on the risky asset ($6\%$); opposite, with exponential salary increase the rate of increase of annual targets is $r^*=7.76\%$ that is larger than the expected return on the risky asset.
\end{itemize}

\section{Changing the retirement age}
In Section \ref{sec:base-case-results} we have set the retirement age equal to 65. It is clear that results strongly depend on the retirement age. In this section we consider different retirement ages, namely 60, 63, 65, 67 and 70, and investigate how the distribution of the final pension $P_{tot}$ changes accordingly.

Table 2 reports, for each retirement age considered $x_T=x$, the annuity value $\ddot{a}_{x}$, the conversion factor from lump sum into pension $\beta_x$, and, for both cases of exponential and linear growths, the old pension $P_o$, the new pension $P_n$, the old net replacement ratio $\Pi_o$ and the new net replacement ratio $\Pi_n$.

\begin{center}
\begin{tabular}{l|l||ll|llll||llll}
\hline
$\mathbf{x_T}$ & $\mathbf{T}$ & $\mathbf{\ddot{a}_{x}}$ & $\boldsymbol{\beta_x}$ & $\mathbf{P_o^e}$ & $\mathbf{P_n^e}$ & $\mathbf{\Pi_o^e}$ & $\mathbf{\Pi_n^e}$ & $\mathbf{P_o^l}$ & $\mathbf{P_n^l}$ & $\mathbf{\Pi_o^l}$ & $\mathbf{\Pi_n^l}$\\ \hline
60 & 30 & 20.95 & 0.048 & 3.63 & 1.57 & 0.6 & 0.26 & 2.04 & 1.26 & 0.6&0.37\\ \hline
63 & 33 & 19.11 & 0.052 & 4.78 & 2.15 & 0.66 & 0.3 & 2.4 & 1.63 & 0.66&0.45\\ \hline
65 & 35 & 17.88 & 0.056 & 5.72 & 2.66 & 0.7 & 0.33 & 2.66 & 1.94 & 0.7&0.51\\ \hline
67 & 37 & 16.64 & 0.06 & 6.81 & 3.29 & 0.74 & 0.36 & 2.93 & 2.3 & 0.74&0.58\\ \hline
70 & 40 & 14.81 & 0.068 & 8.82 & 4.56 & 0.8 & 0.41 & 3.36 & 2.98 & 0.8&0.71\\ \hline
\end{tabular}\vspace{0.3cm}\\
Table 2.
\end{center}

\begin{figure}[H]
		\centering
\includegraphics[width=.36\columnwidth]{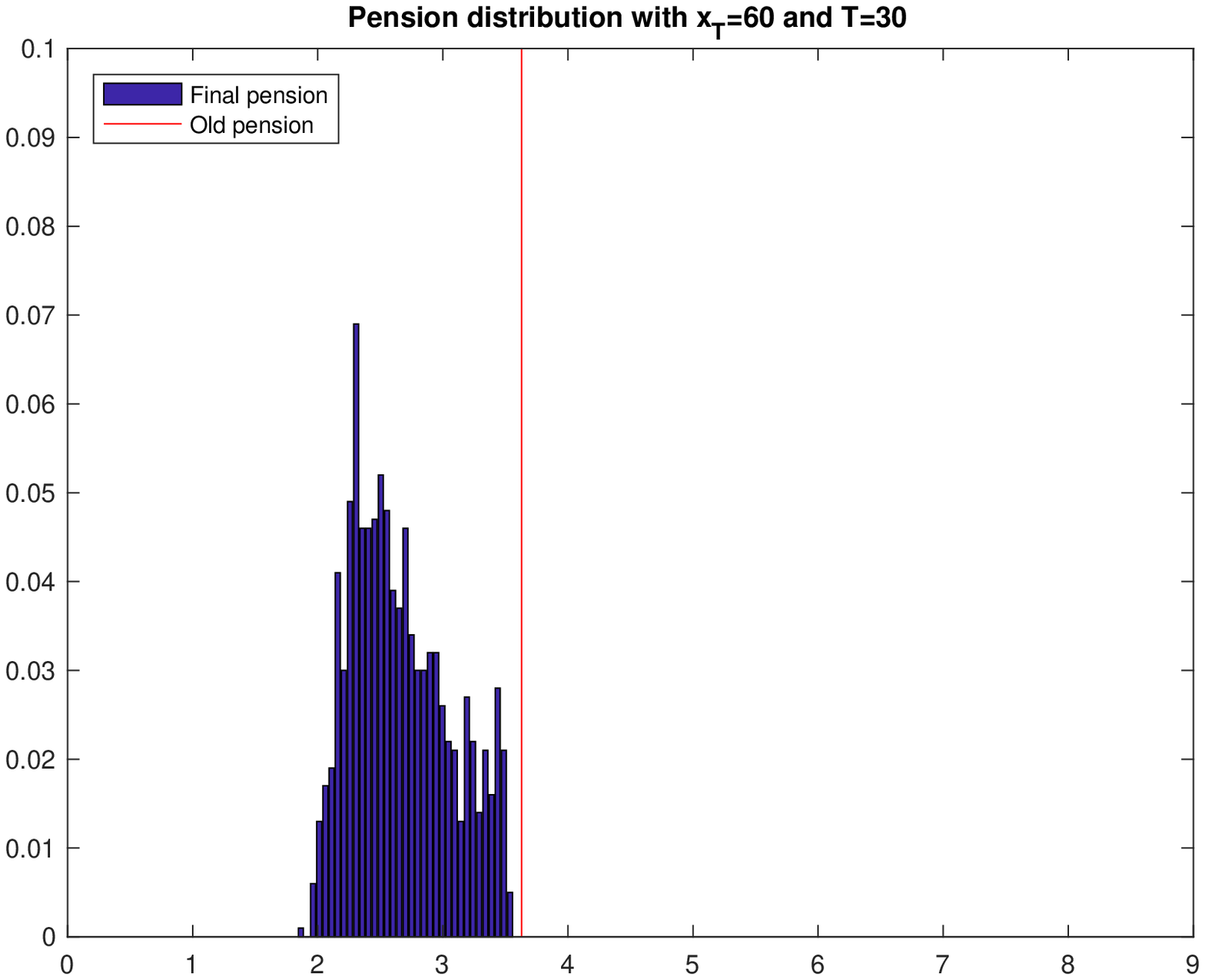}\includegraphics[width=.36\columnwidth]{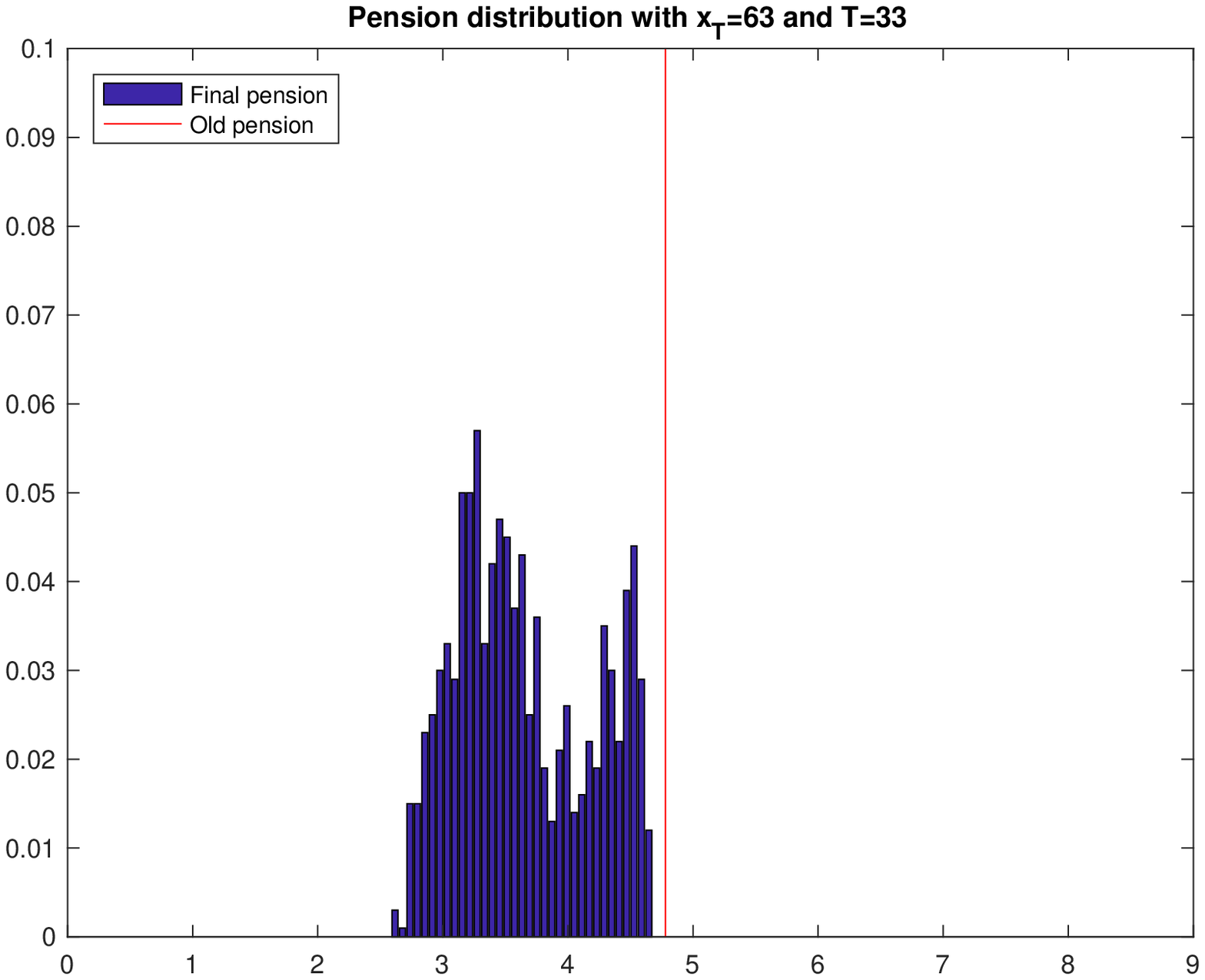}\includegraphics[width=.36\columnwidth]{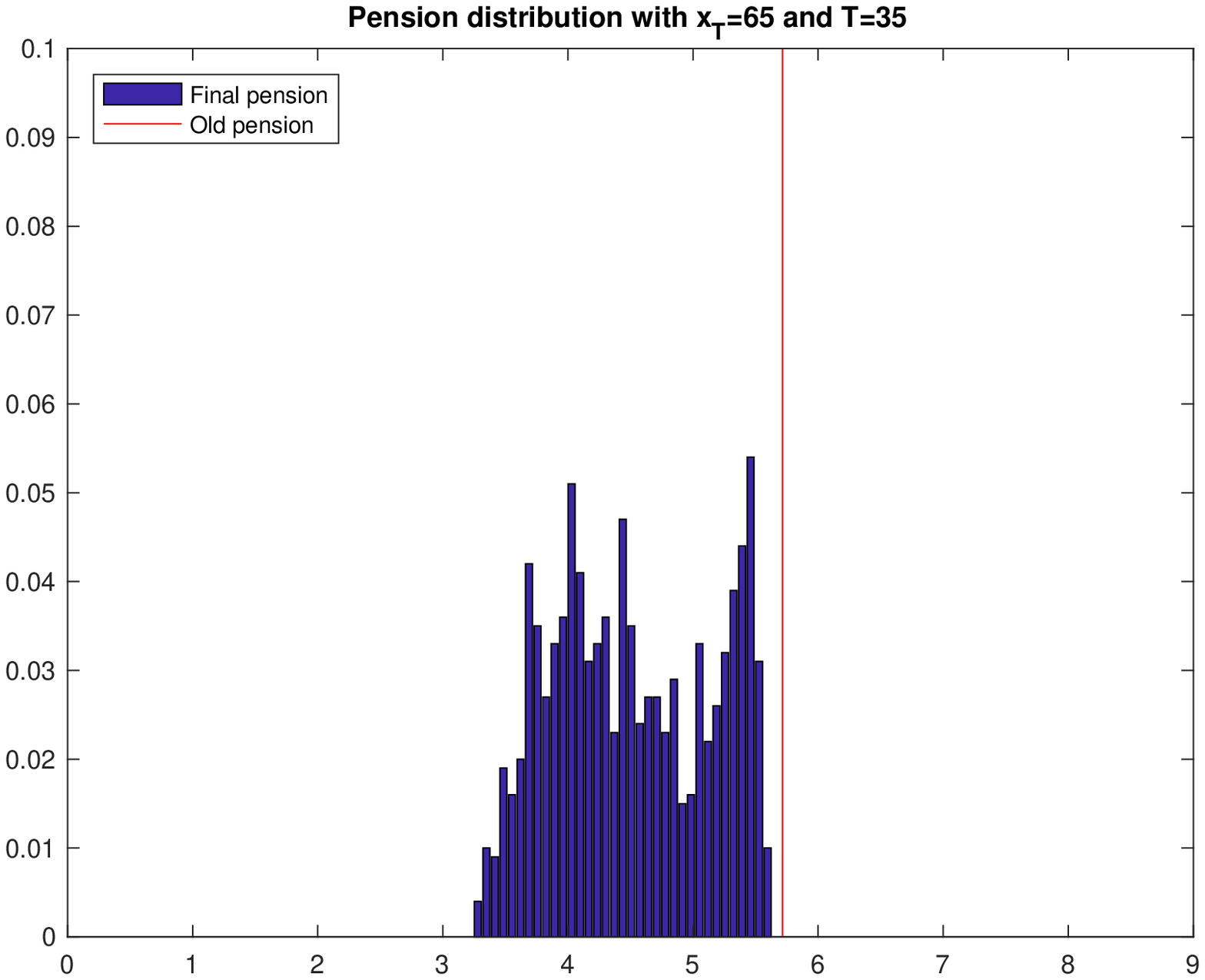}
\includegraphics[width=.36\columnwidth]{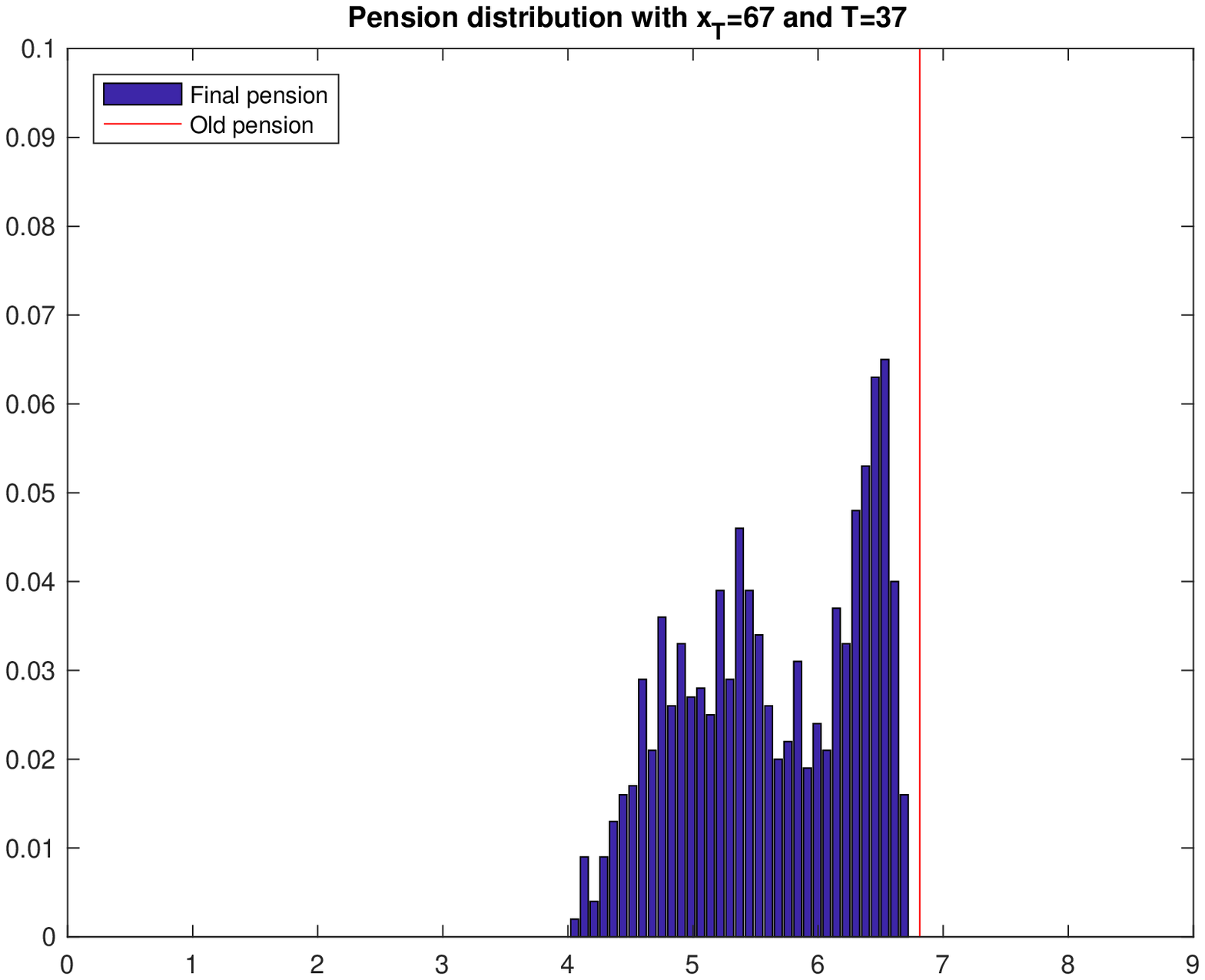}\includegraphics[width=.36\columnwidth]{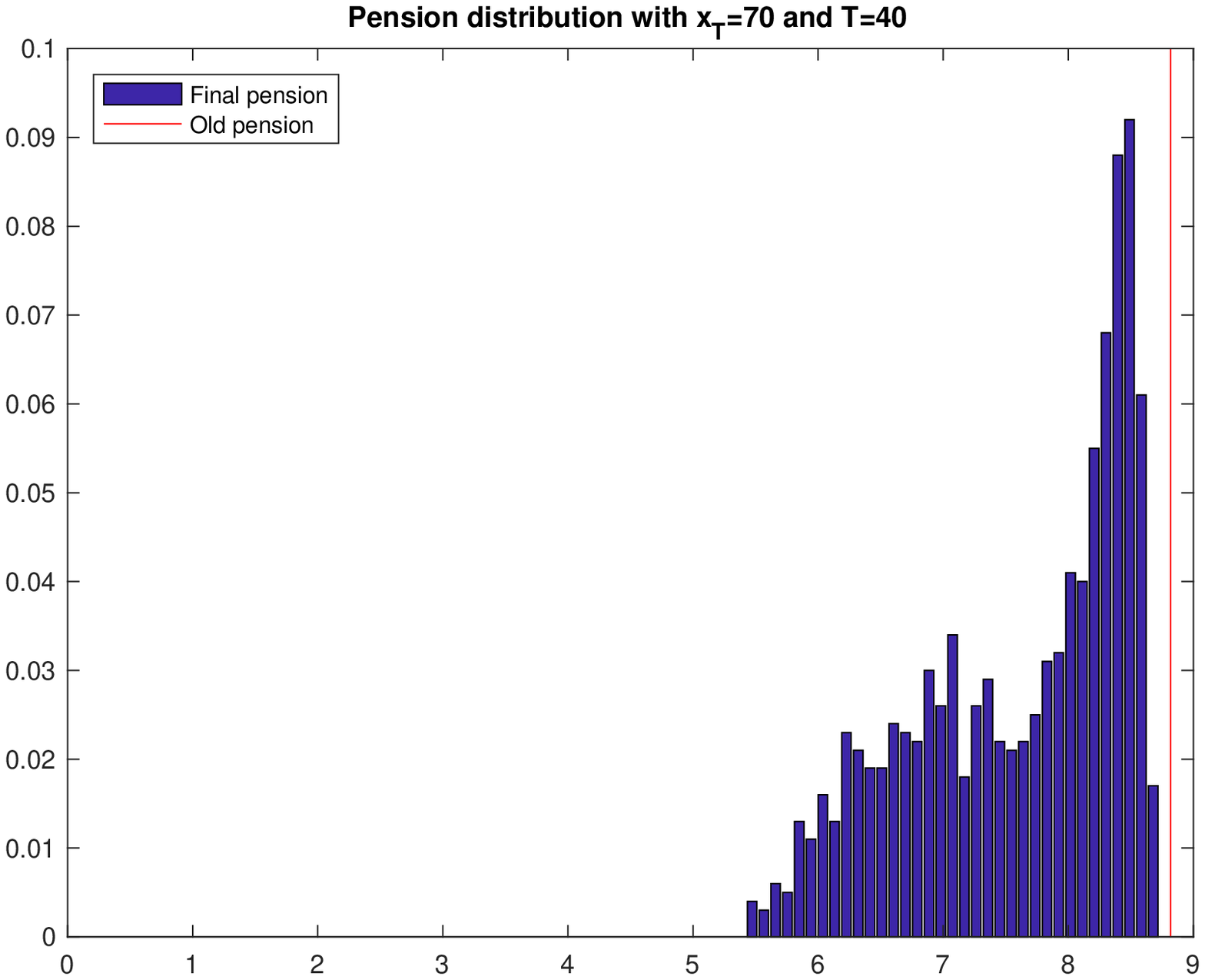}
\caption{Pension distribution with different retirement ages (exponential growth)}
\label{fig:age}
\end{figure}

\begin{figure}[H]
		\centering
\includegraphics[width=.36\columnwidth]{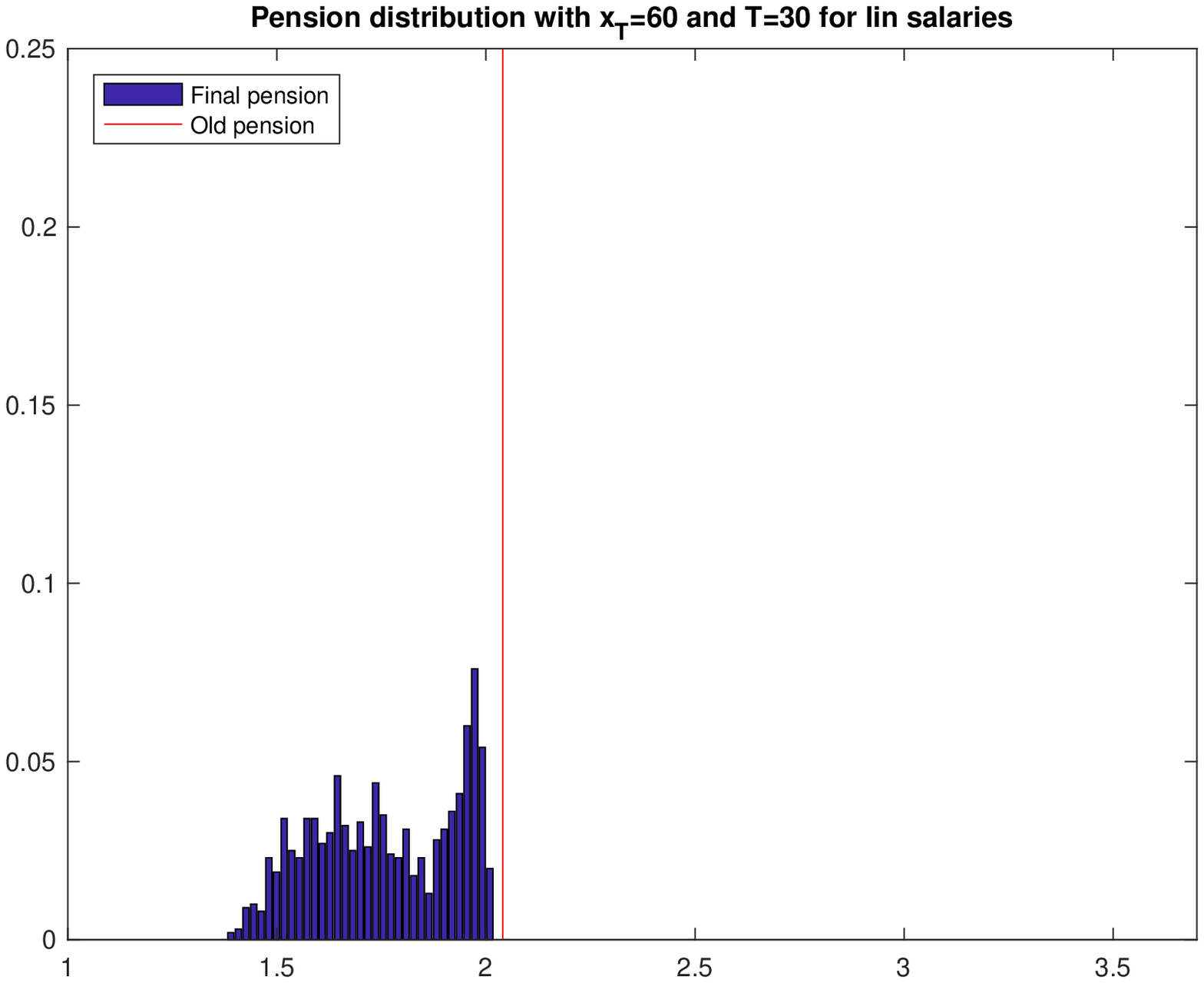}\includegraphics[width=.36\columnwidth]{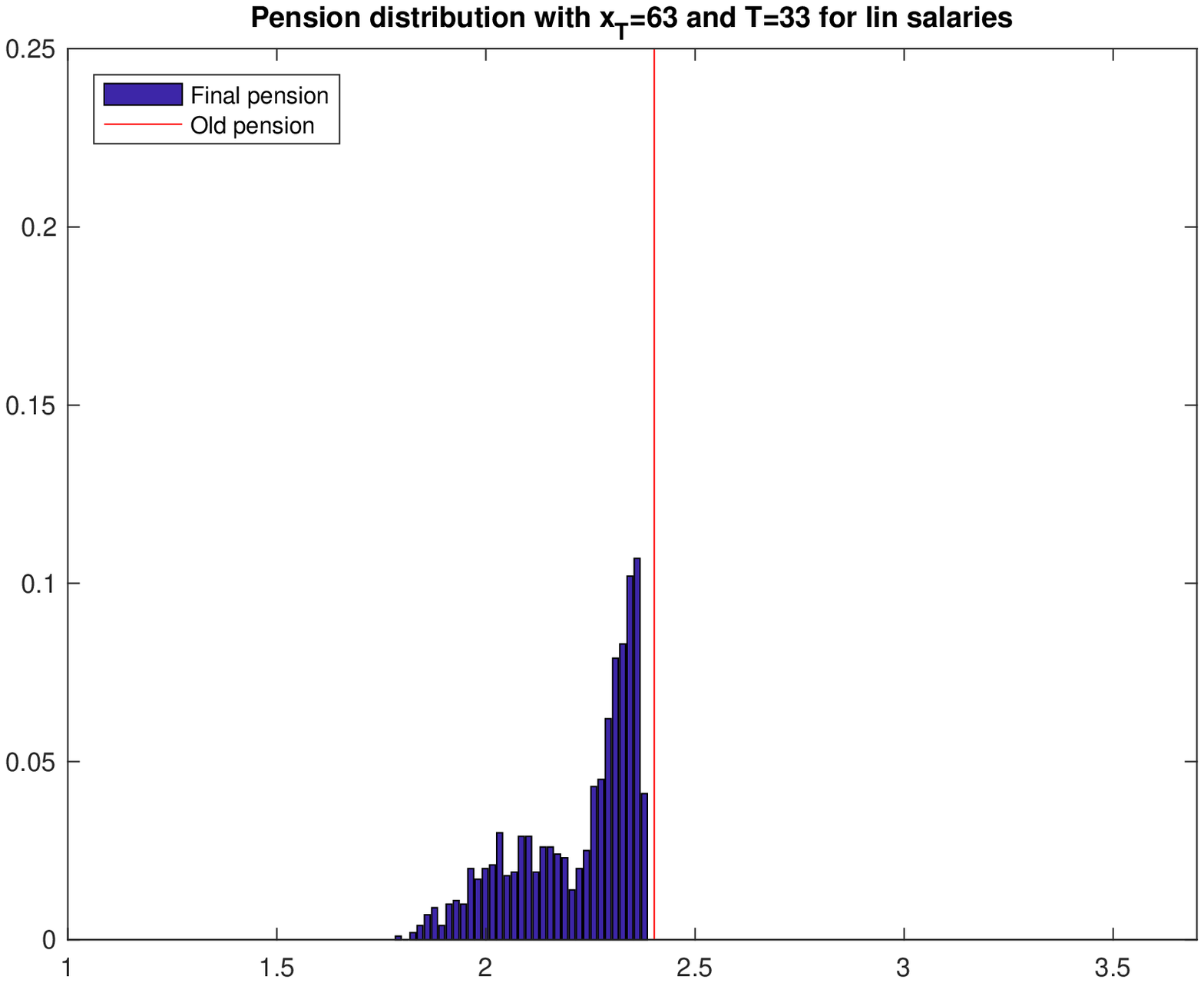}\includegraphics[width=.36\columnwidth]{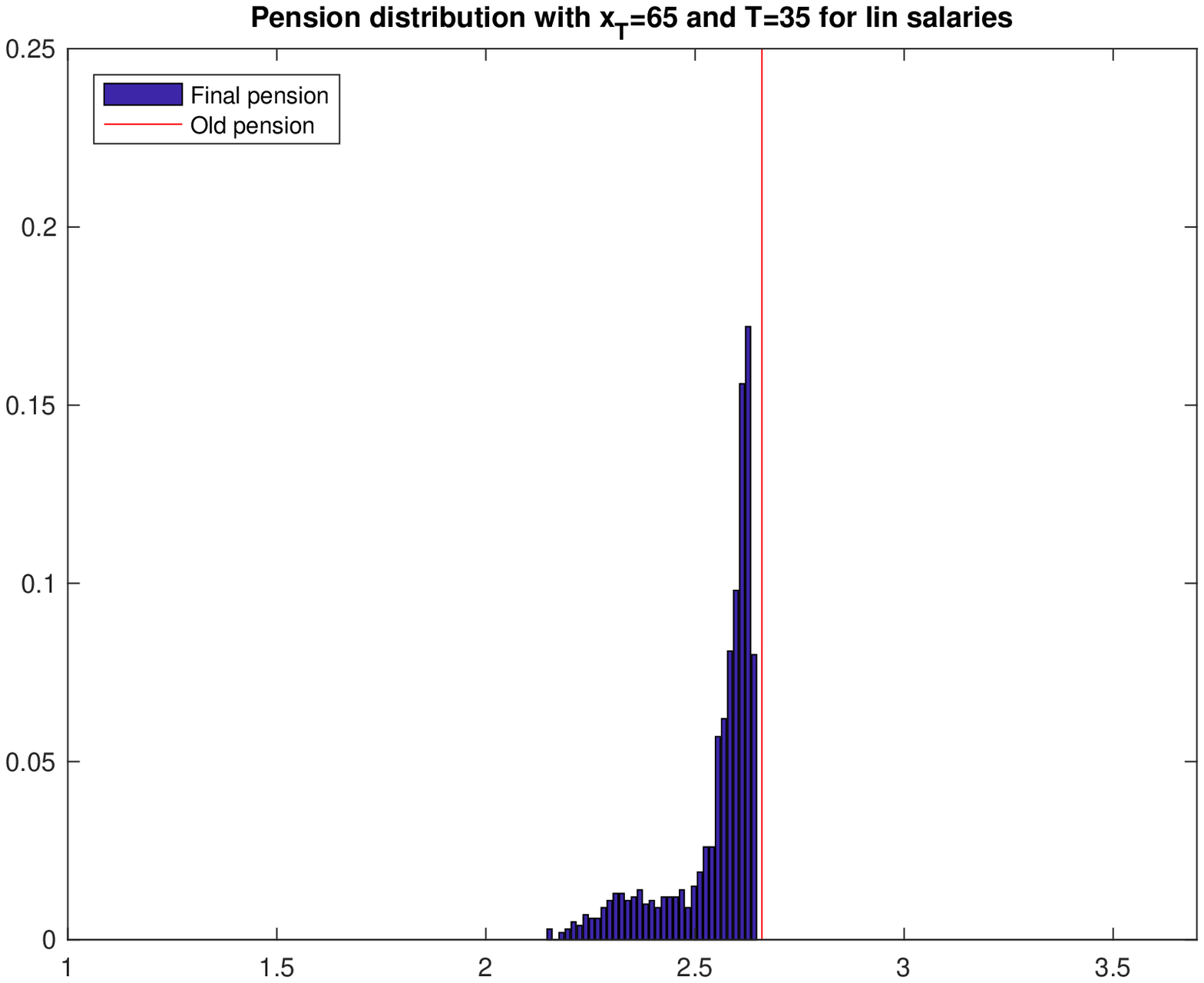}
\includegraphics[width=.36\columnwidth]{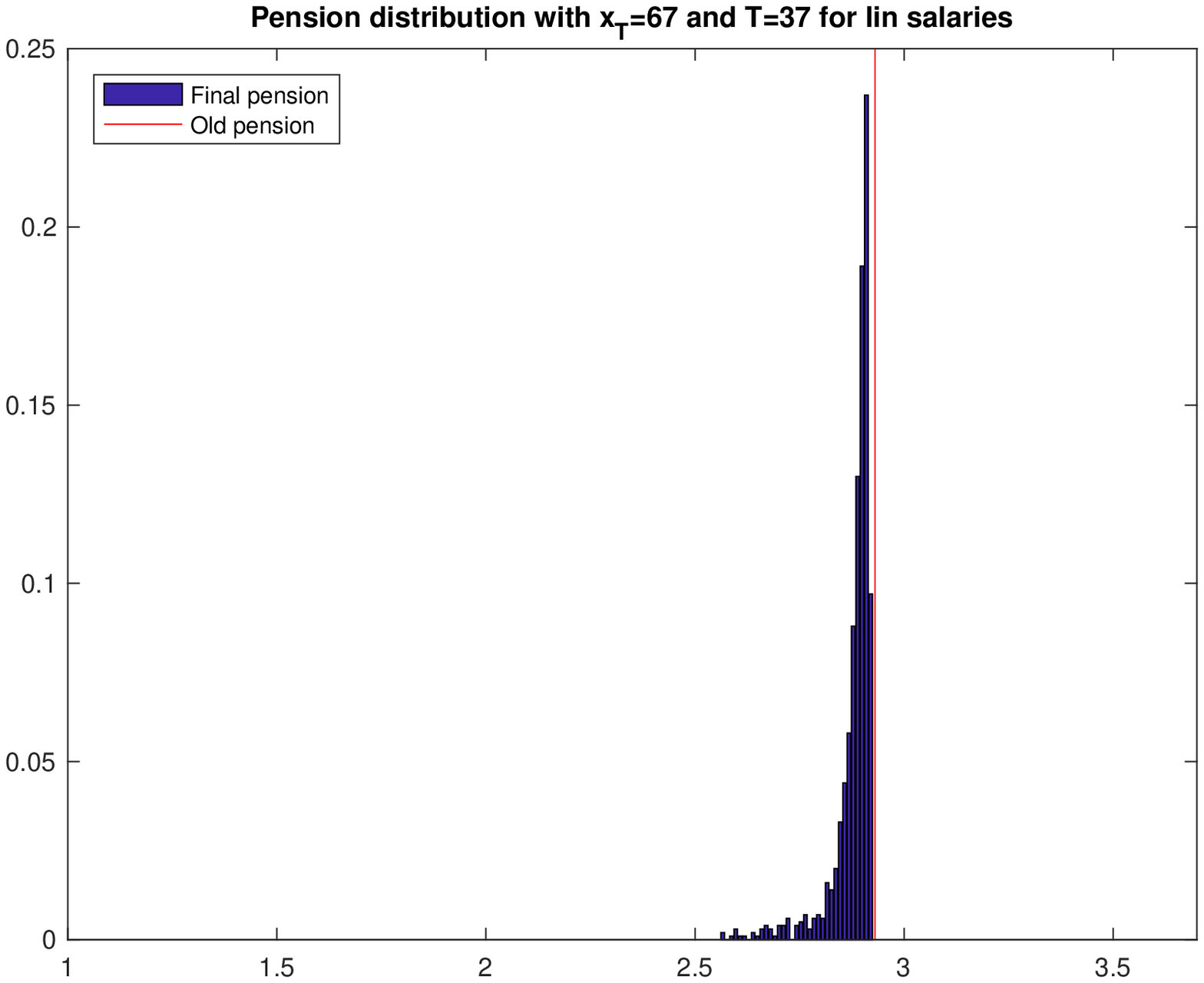}\includegraphics[width=.36\columnwidth]{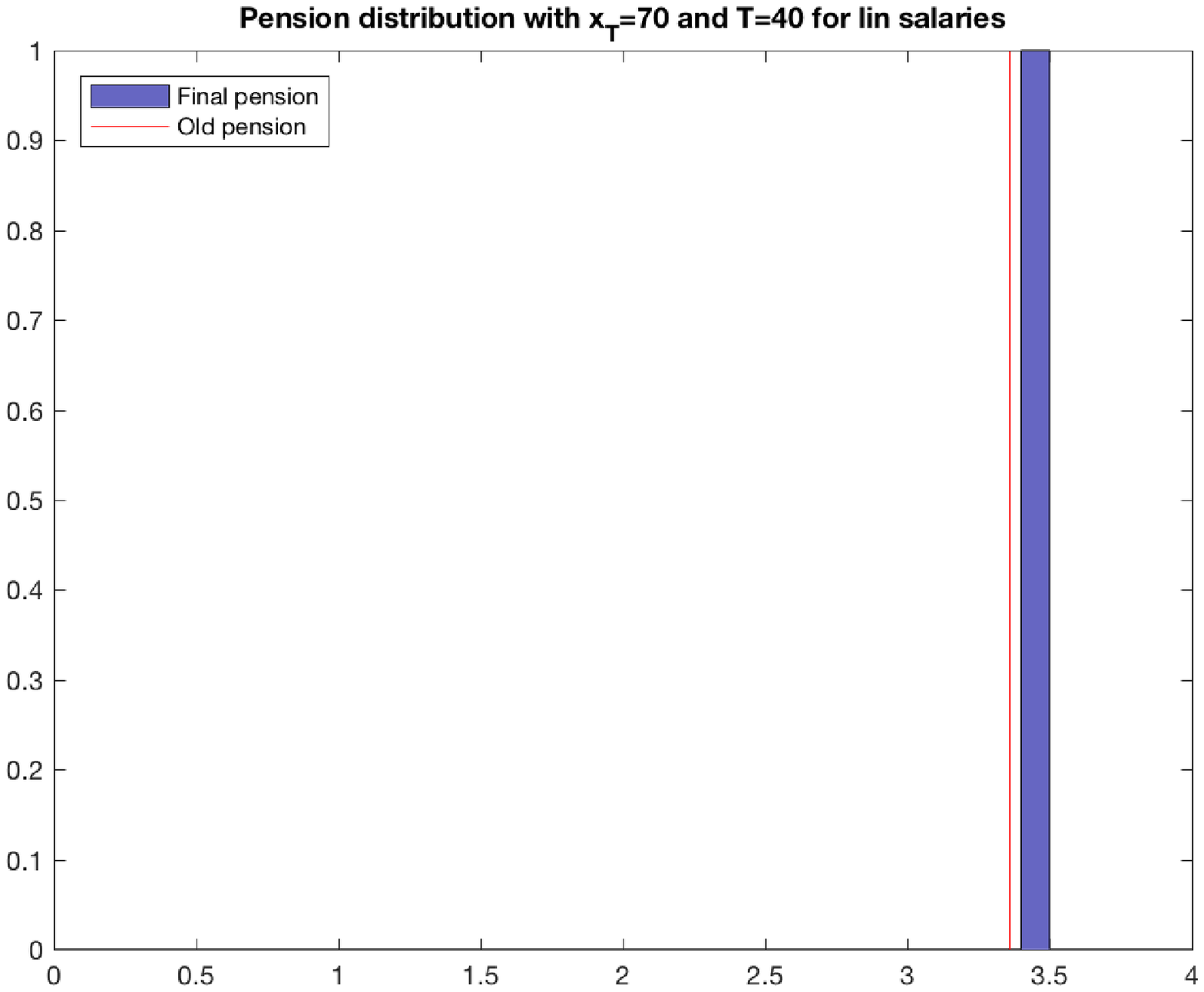}
\caption{Pension distribution with different retirement ages (linear growth)}
\label{fig:age-lin}
\end{figure}

The graphs in Figure \ref{fig:age} report the pension distribution with the different retirement ages considered in the case of exponential salary growth, while the graphs in Figure \ref{fig:age-lin} report those related to the linear salary increase.

We notice the following:
\begin{itemize}
\item The comparison between exponential and linear salary increase confirms what already observed in Section \ref{sec:base-case-results} at all ages: the distribution of the final pension is more spread out in the area on the left of the old pension in the case of exponential salary increase, while it is more peaked immediately on the left of the old pension in case of linear target, showing the larger chance of approaching the target in the case of linear increase.
\item With both salary increases, we observe that it is easier to reach the target with an older retirement age: higher retirement age means lower gap between the old and the total pension. This is intuitive and expected, and is due to different reasons: (i) because of actuarial fairness principles, the highest the retirement age, the lower the price of the lifetime annuity; (ii) a higher retirement age also means that the fund grows for a longer time, meaning a higher lump sum to be converted into pension. These two factors imply that the higher the retirement age, the higher the final pension, everything else being equal.
\item In the extreme case of linear salary increase and retirement age equal to 70, the difference between the old and the new pension is so small that investing in the riskless asset for the entire working life (40 years) is sufficient to cover the gap, and the final pension $P_{tot}=3.464$ turns out to be higher than the old pension $P_o=3.36$ in $100\%$ of the cases.
\end{itemize}

\section{Break even points }
\begin{figure}[H]\label{fig:break-even-points}
		\centering
\subfloat[][Break even point for $\b$ (exp salary)]
{\includegraphics[width=.42\columnwidth]{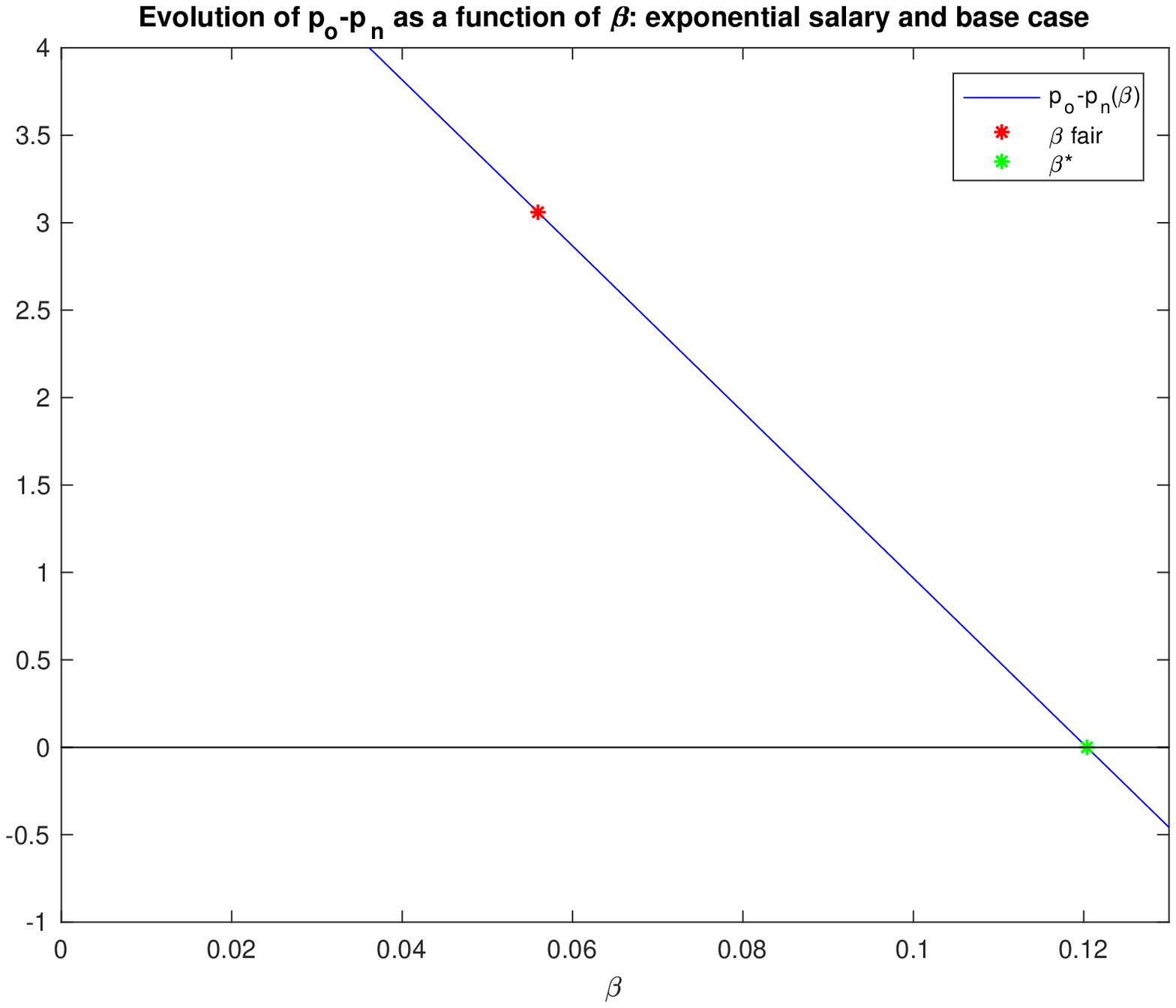}\label{bep-exp-beta}}
\qquad \qquad
\subfloat[][Break even point for $\b$ (lin salary)]
{\includegraphics[width=.42\columnwidth]{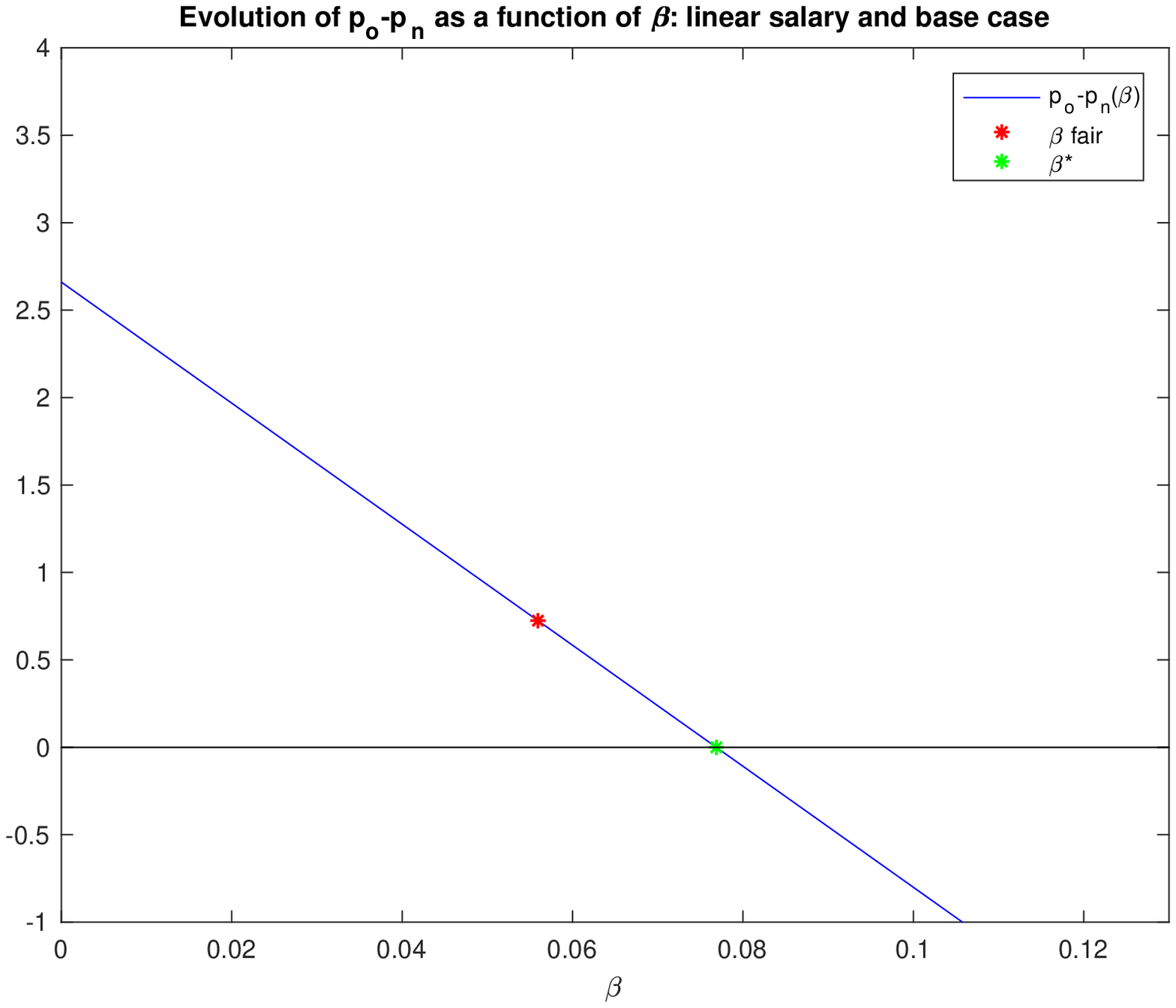}\label{bep-lin-beta}}
\\		
\subfloat[][Break even point for $w$ (exp salary)]
{\includegraphics[width=.42\columnwidth]{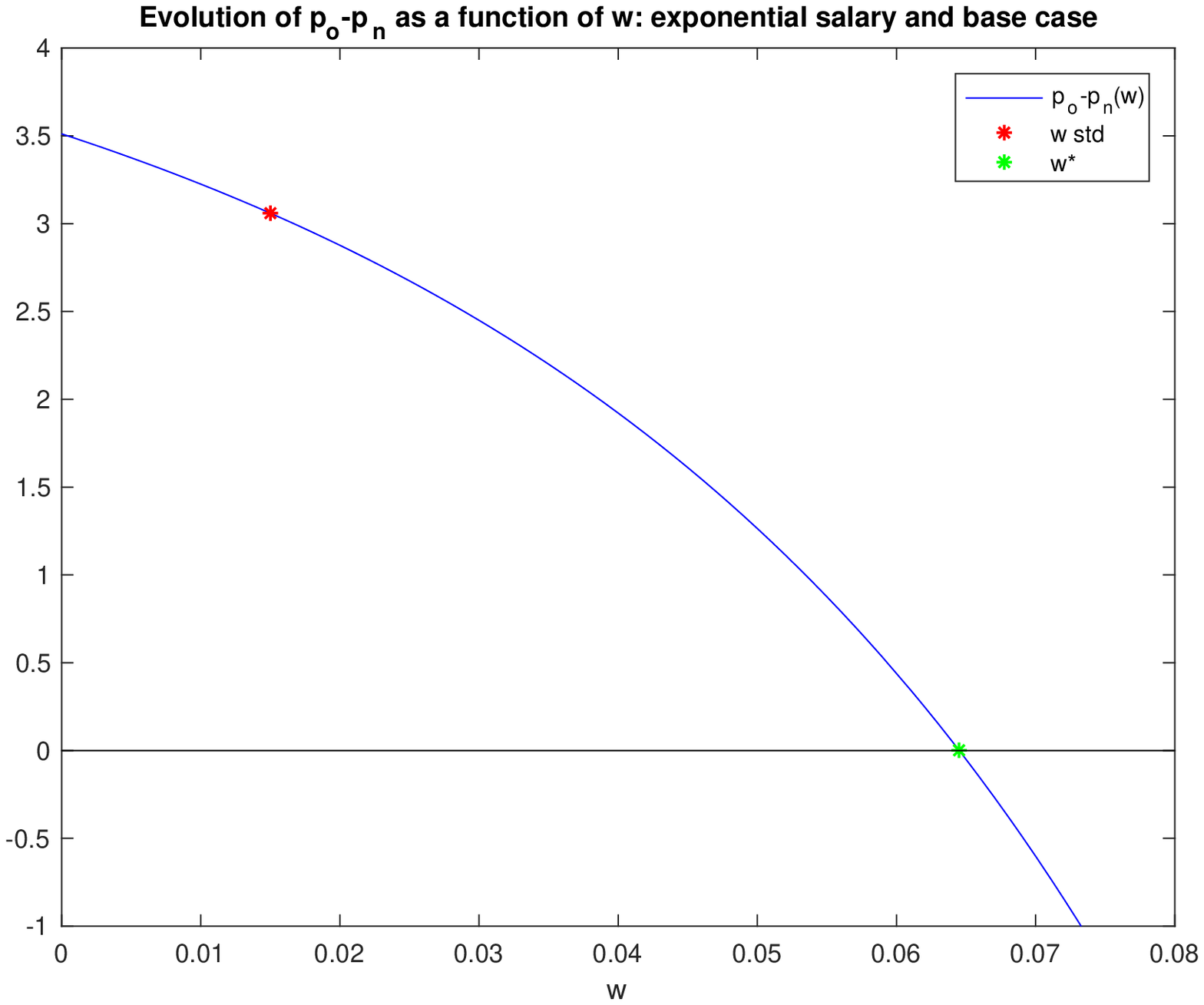}\label{bep-exp-w}}
\qquad \qquad
\subfloat[][Break even point for $w$ (lin salary)]
{\includegraphics[width=.42\columnwidth]{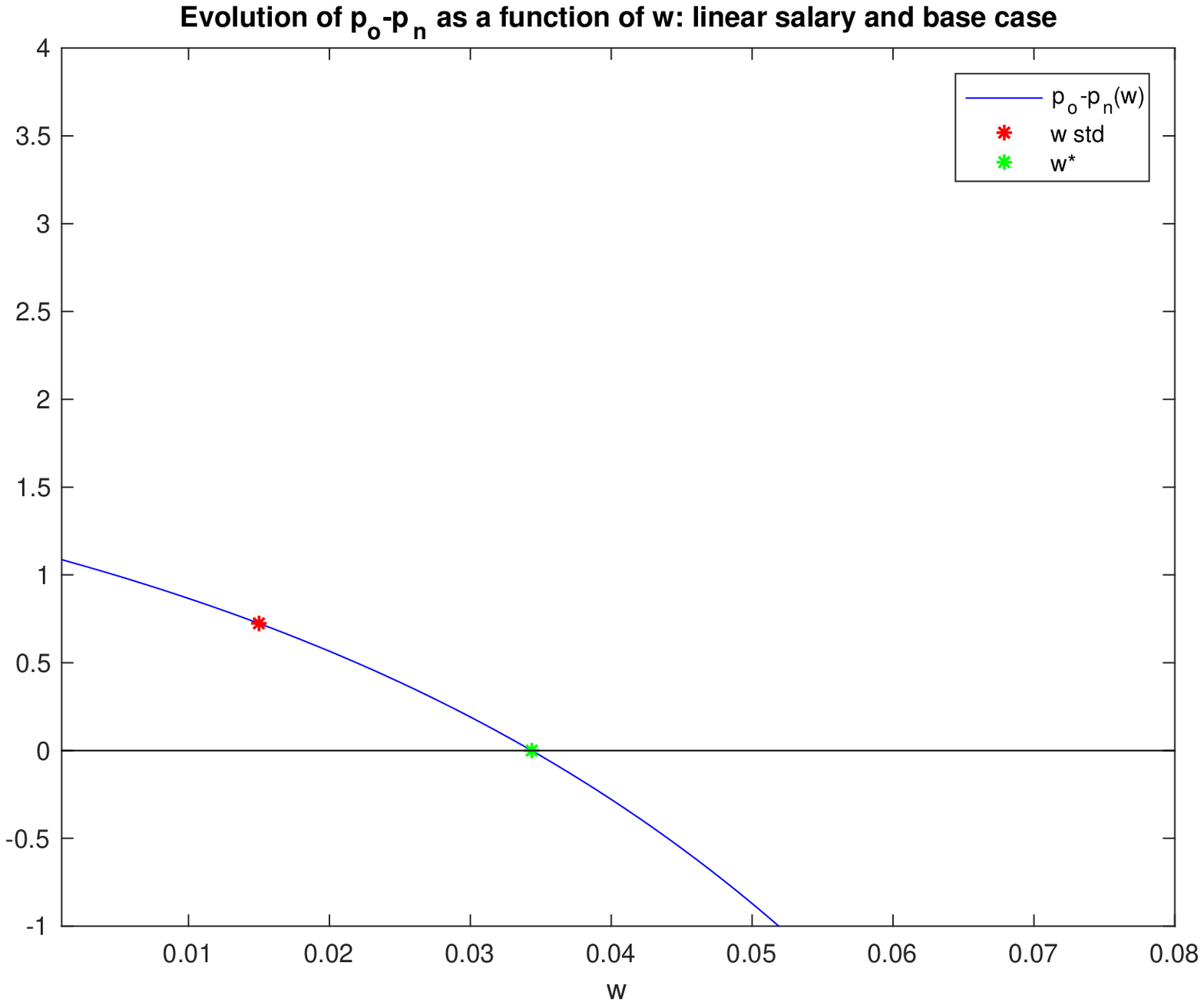}\label{bep-lin-w}}
\\
\subfloat[][Break even point for $g$ (exp salary)]
{\includegraphics[width=.42\columnwidth]{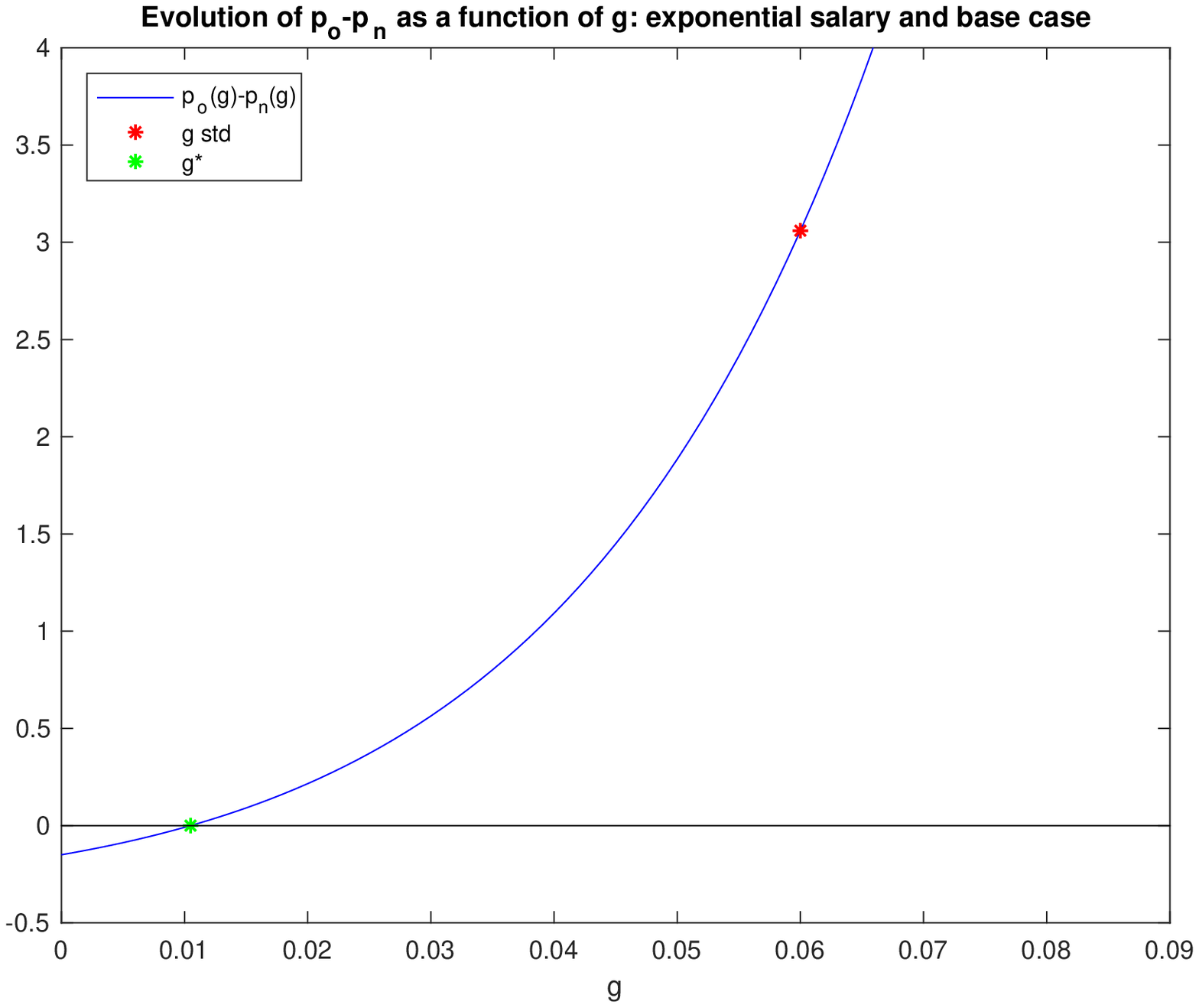}\label{bep-exp-g}}
\qquad \qquad
\subfloat[][Break even point for $g$ (lin salary)]
{\includegraphics[width=.42\columnwidth]{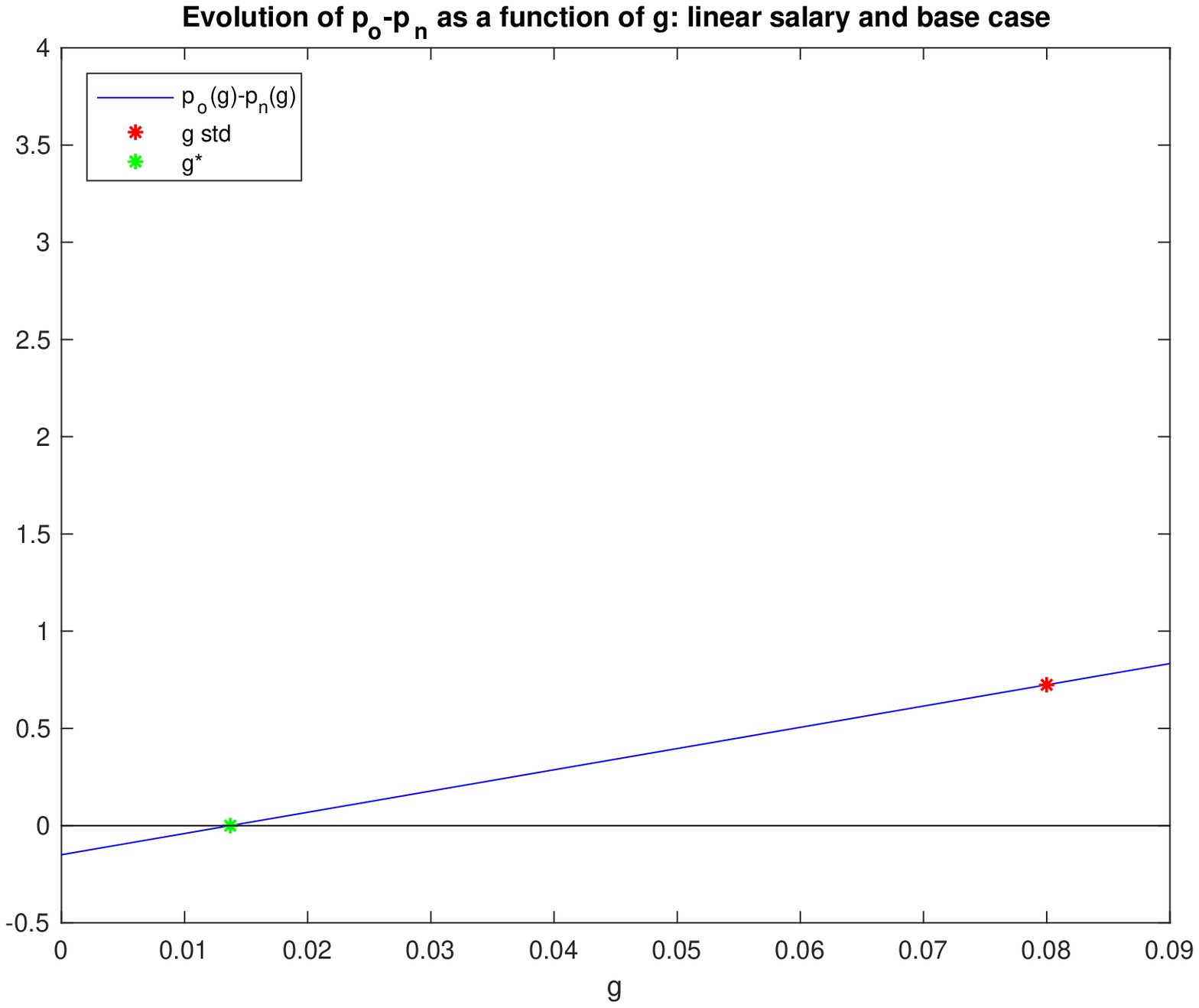}\label{bep-lin-g}}
\caption{Break even points w.r.t. $\b$, $g$ and $w$ for exponential and linear salary.}
	\end{figure}

In the previous sections we have seen that, expectedly, with both salary growths the old pension is larger than the new pension. This result heavily depends on the choice of the parameters and may no longer hold if some parameters change. We have calculated what values of some key parameters would equate the old pension to the new pension, leaving the values of the remaining parameters equal to those of the base case. In particular, we have calculated the break even points for the parameters $w$, $\b$ and $g$. Figure 9 reports the six plots of the quantity $P_o-P_n$ (difference between the old and the new pension) as a function of $\b$, $g$ and $w$, in both cases of exponential and linear salary. In particular, figures \ref{bep-exp-beta}, \ref{bep-exp-g} and \ref{bep-exp-w} report the exponential salary case, while figures \ref{bep-lin-beta}, \ref{bep-lin-g} and \ref{bep-lin-w} report the linear salary case. The red point is the base case, the green point on the $x-$axis is the break even point that equals the old pension to the new pension.

We notice the following:
\begin{itemize}
\item the difference between the old and the new pension decreases with $\b$, i.e., it increases with the price of the annuity $1/\b$. This is obvious, because the old pension is not affected by the price of the annuity, while the new pension is affected by $\b$ and it increases with it; therefore, the higher $\b$ the higher $P_n$, the lower $P_o-P_n$. With exponential increase, the old and the new pension are equal when $\b=0.12$, that corresponds to price of the unitary annuity equal to approximately 8.33, against the base value of 17.785; with linear increase, the old and the new pension are equal when approximately $\b=0.078$, that corresponds to price of the unitary annuity equal to 12.82, against the base value of 17.785;

       \item the difference between the old and the new pension decreases with $w$. This is again obvious, because the old pension is not affected by the mean GDP growth rate $w$, while the new pension is affected by $w$ and it increases with it; therefore, the larger $w$ the larger the new pension, the lower the gap between the old and the new pension. With exponential increase, the old and the new pension are equal with a mean GDP of approximately $w=6.5\%$, with linear increase, the old and the new pension are equal with mean GDP of approximately $w=3.5\%$;

           \item the difference between the old and the new pension increases with $g$. This result is interesting, because both the old pension and the new pension are positively correlated with the salary growth $g$, but to a different extent: the old pension is affected by it only via the final salary that is used to calculate the pension income, the new pension is affected by it via the yearly contributions that are paid into the fund and accumulated until retirement. Figures \ref{bep-exp-g} and \ref{bep-lin-g} seem to suggest that the impact of $g$ on the old pension is larger than that on the new pension, leading to a larger gap in case of increase of $g$;

               \item the break even point for the salary increase rate is about $g_e=1\%$ for exponential salary increase, about $g_l=1.5\%$ for linear increase. This result indicates that with sufficiently small salary increase the old pension and the new pension coincide. In the presence of salary increase rates smaller than the break even point, the new pension is larger than the old one. This is consistent with what observed in point 3. in Section \ref{sec:base-case-results}: the effect of the pension reform is more considerable for workers with dynamic career than for workers with smooth career.
\end{itemize}

\section{Conclusions}
In this paper, we have tackled the issue of the gap between an ``old'' pre-reform salary-related pension and a ``new'' post-reform contribution-based pension. We have investigated to what extent the gap can be reduced by adding to the state pension another pension provided by a DC pension scheme. We have used stochastic optimal control and a target-based approach to find the optimal investment strategy suitable to cover the gap between the salary-related and the contribution-based pension. The numerical simulations suggest that the gap between the salary-related pension and the contribution-based pension is larger for workers with dynamic career than for workers with a stagnant career, meaning more difficulties for the first class of workers to fill in the gap than for the second class of workers, even by assuming a higher savings capacity. Intuitively, the gap is easier to cover in the case of late retirement, and vice versa. This result is consistent with results in \citeasnoun{borella-codamoscarola-jpef}. A slow salary increase associate to late retirement age can produce a new pension that is almost equal to (or even exceeds) the old pension. Expectedly, the gap reduces when the mean GDP increase and when the price of the annuity increases. Interestingly, the gap increases when the rate of increase of the salary increases.

\bibliographystyle{dcu}
\bibliography{file-bib}

\end{document}